\documentclass{aa}  

\usepackage{graphicx}
\usepackage[varg]{txfonts}  
\usepackage{natbib} 
\bibpunct{(}{)}{;}{a}{}{,} 

\newcommand{\MSun}{M_\mathrm{\odot}}
\newcommand{\SgrAstar}{\protect{Sgr~A*\,\,}}

\begin{document} 

   \title{How to create Sgr A East}
   
  \subtitle{Where did the supernova explode?}
  
   \author{S. Ehlerov\' a
          \inst{1}
          \and
          J. Palou\v s
          \inst{1}
          \and 
          M. R. Morris
          \inst{2}
          \and
          R. W\" unsch
          \inst{1}
          \and 
          B. Barna
          \inst{3}
          \and 
          P. Vermot
          \inst{1} 
          }
     \institute{Astronomical Institute of the Czech Academy of Sciences,
              Bo\v{c}n\'{\i} II 1401/1, 140 00 Prague\\
              \email{sona.ehlerova@asu.cas.cz}
              \and
           Department of Physics and Astronomy, University of California, Los Angeles, CA 90095-1547, USA
              \and
           Physics Institute, University of Szeged, D\'{o}m t\'{e}r 9, Szeged, 6723, Hungary
             }

   \date{Received 4 August 2022; accepted 27 October 2022}

 
  \abstract
   {Sgr A East is the supernova remnant closest to the centre of the Milky Way. Its age has been estimated to be either very young, around 1-2 kyr, or about 10 kyr, and its exact origin remains unclear.}
   {We aspire to create a simple model of a supernova explosion that reproduces the shape, size, and location of Sgr A East.}
   {Using a simplified hydrodynamical code, we simulated the evolution of a supernova remnant in the 
   medium around the Galactic centre. The latter consists of a nearby massive molecular cloud with which Sgr A East is known to be interacting and a wind from the nuclear star cluster.}
   {Our preferred models of the Sgr A East remnant are compatible with an age of around 10 kyr.
   We also find suitable solutions for older ages, but not for ages younger than 5 kyr. 
   Our simulations predict that the supernova exploded at a distance of about 3.5 pc from the Galactic centre, below the Galactic plane, slightly eastwards from the centre and 3 pc behind it.}
   {}


   \maketitle
%

\section{Introduction}

\begin{figure}
   \includegraphics[angle=0,width=0.45\textwidth]{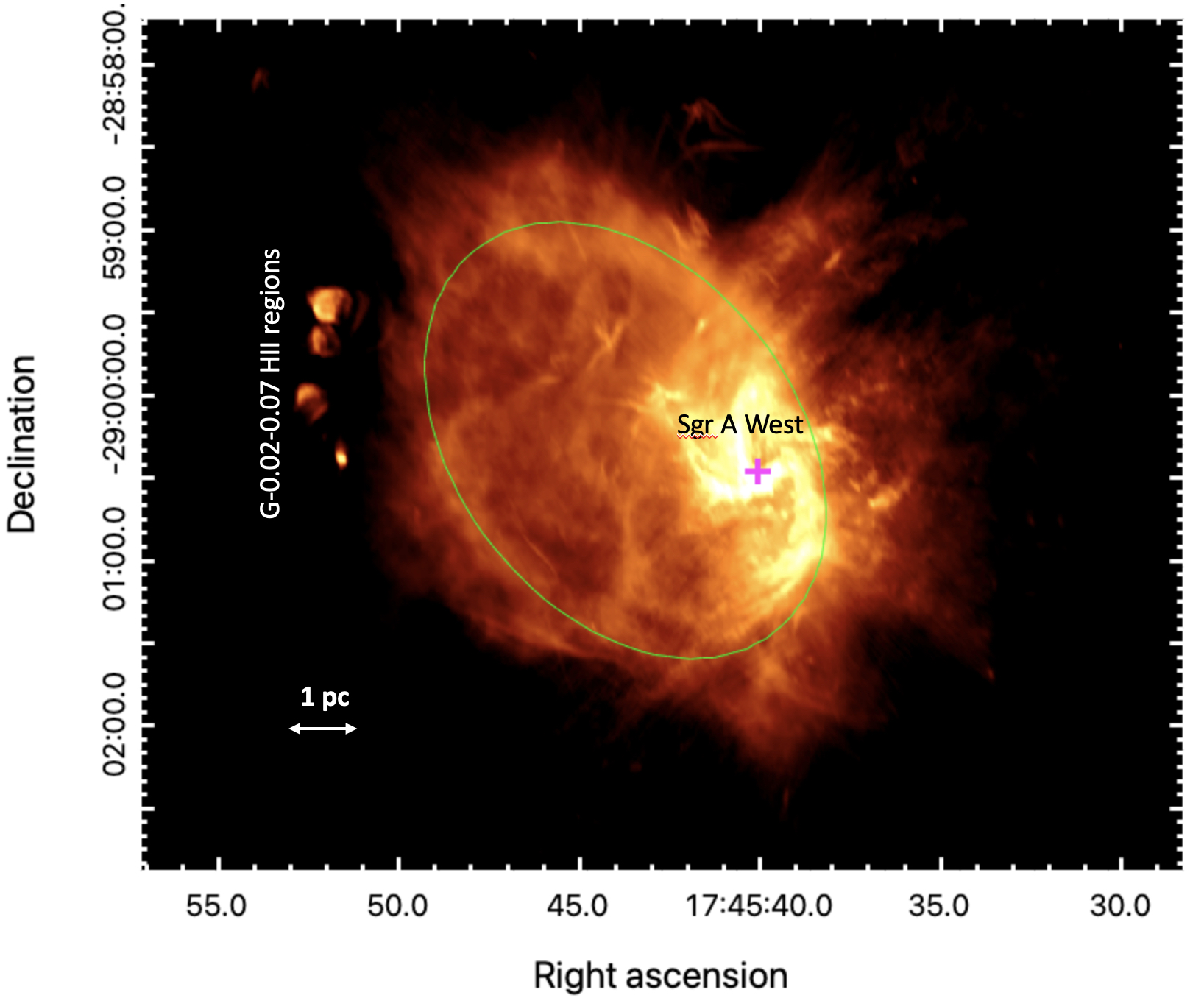}
      \caption{Radio image of the Sgr A East supernova remnant at 6 cm wavelength \citep{2016ApJ...817..171Z}. Coordinates are J2000.  The thin green ellipse shows the approximate outline of the expanding shell.  The location of \SgrAstar is marked with a magenta plus (+).   The bright Sgr A West HII region surrounding \SgrAstar is in the near foreground of Sgr A East. The compact HII regions constituting G-0.02-0.07, appearing at the left, are embedded in the 50 $\mbox{kms}^{-1}$ molecular cloud with which Sgr A East is interacting.
              }
         \label{SgrAE}
\end{figure}

Sagittarius A East (Sgr A East) is the supernova remnant closest to \SgrAstar, the supermassive black hole in the centre of the Milky Way. The morphology and extent of Sgr A East are well displayed by low-frequency radio images, such as the 6cm images shown in Fig. \ref{SgrAE}. Sgr A East belongs to the class of radio-bright supernova remnants; it is the 18th brightest remnant in the Milky Way out of 194 in \cite{green2019}.

The projected distance between the centre of Sgr A East and \SgrAstar is $\approx 1'$, which corresponds to 2.3 pc assuming the distance of Sgr A* to be 8200 pc \citep{2019A&A...625L..10G}.
The dimensions of Sgr A East are about $3.5' \times 2.5'$ (= $8.3 \times 6.0$ pc)
or perhaps slightly larger (the ellipse in Fig. \ref{SgrAE} corresponds to the cited values).
On its eastern side, Sgr A East is clearly interacting with the 50 $\mbox{kms}^{-1}$ giant molecular cloud \citep[see][]{2020ApJS..249...35B,1992ApJ...395..166S}.

Sgr A East is a mixed-morphology supernova remnant, that is, it has a shell-like 
appearance in radio and is centrally bright in X-rays \citep[for other examples see,
e.g.][]{rho1998}. Mixed-morphology supernova remnants are usually associated with denser-than-average surroundings. They often interact with molecular clouds or generally expand into a complex surrounding (e.g. model of IC 443 in \citet{2021A&A...649A..14U} and many others).  They are usually considered middle-aged: recent papers about mixed-morphology supernova remnants derive ages between 
5 to 20 kyr. 

Close to the north-northeastern rim of Sgr A East, there is a pulsar wind nebula called the Cannonball, which is detected both in radio and X-ray emission \citep{2005ApJ...631..964P,zhao2013}. This object is commonly considered as the candidate stellar remnant from the event that produced Sgr A East, but this suggestion has been disputed \citep{yalinewich2017}. 

Even though \SgrAstar and the surrounding HII region (Sgr A West, with its so-called minispiral) are projected onto the interior of Sgr A East, it is estimated that the supernova remnant lies
behind \SgrAstar because of the absorption of the low-frequency nonthermal radio emission from Sgr A East by the thermal gas in Sgr A West \citep{1987AJ.....94.1178Y,1989ApJ...342..769P}.
Based on an analysis of masers connected to Sgr A East (possible interaction zone between Sgr A East and the molecular cloud), its 3D distance from the Galactic centre would be around 5 pc \citep{1999ApJ...512..230Y}.

Sgr A East is usually considered to have originated from a standard Type II supernova,
but there are also theories connecting it to the tidal disruption of a star by
the supermassive black hole \citep{KhokhlovMelia96,guillochon2016} or to a rare Type Iax supernova
\citep{zhou2021}. In the following, we assume a standard Type II supernova origin.

The age of Sgr A East is unknown, but it is estimated to be between 1 and 10 kyr.
Broadly speaking, there are two groups of age estimates, one advocating an age of about 10 kyr,
and the other preferring the age of about 2 kyr. One of the main differences between these groups
is that the greater age is compatible with the neutron star Cannonball being connected
to the explosion of the supernova that created Sgr A East, while the younger age
is not, in which case the Cannonball is unrelated.

\cite{mezger1989}, in their paper about the circumnuclear disk, modelled Sgr A East
as an explosion in a medium having an ambient density of $10^4\mbox{cm}^{-3}$ with an energy of $4\times10^{52}\, \mbox{erg}$
(= 40 supernovae, with $E_{\mathrm{SN}}=10^{51}\, \mbox{erg})$, and an age of 7.5 kyr.  No other models invoking such a high energy have since appeared.  

\cite{maeda2002} gave a thorough description of Sgr A East, both in radio (VLA) 
and X-ray (Chandra) emission, and they provided a schematic picture of Sgr A East (their Fig. 1).
Maeda et al. described Sgr A East as a metal rich, compact, mixed-morphology supernova
remnant, and they described its relative compactness to a very high ambient density in this region. 
They estimated its age at 10 kyr, consistent with the evolution of a standard supernova
into an ambient density of $10^3 \mbox{cm}^{-3}$ and sheared by Galactic rotation.

The putative pulsar, the Cannonball, was identified in X-ray \citep{2005ApJ...631..964P} and in
radio \citep{zhao2013}, where its proper motion was also measured.
The transverse velocity of the Cannonball is $500 \pm 100\ \mbox{kms}^{-1}$, and its direction of motion is compatible with having started at the explosion site of the supernova which created Sgr A East. In this case, the explosion took place $9\pm2$ kyr ago.

Using mid-infrared observations with SOFIA, \cite{lau2015} detected a patch of dust lying toward the centre of the remnant.  According to their analysis, this dust was created from the supernova ejecta, and survived the passage of the reverse shock some time ago,
though a larger amount of (newly condensed) dust from the explosion was presumably destroyed.

Numerical smoothed-particle hydrodynamics (SPH) simulations by \cite{rockefeller2005} predict a young age of  1700 yr
for Sgr A East; in their model, the supernova explosion evolves in a medium created 
by the winds from the massive stars in the young nuclear star cluster 
and encounters the circumnuclear disk at a distance of 2 - 3 pc from \SgrAstar. 
Simulations using the infinitesimally thin-shell Kompaneets approximation in a power-law
density distribution \citep{rimoldi2015} predict the X-ray observability of supernova
remnants in the vicinity of \SgrAstar to be $\sim 10^4$ yr, which is compatible with the age of Sgr A East.
They, however, predicted a rather spherical shape for evolving remnants, which is not fully
compatible with observations of Sgr A East.
Other simulations, analytical and hydrodynamical, based on the evolution of the remnant 
in a medium dominated by the wind, prefer a young age of 1500 yr \citep{yalinewich2017}.
These authors also predict that supernova remnants around \SgrAstar will not be observable after
about 9 kyr.

\section{Region within 20 pc of \SgrAstar}
\label{sec:inside20pc}

\begin{figure}
   \includegraphics[angle=0,width=0.45\textwidth]{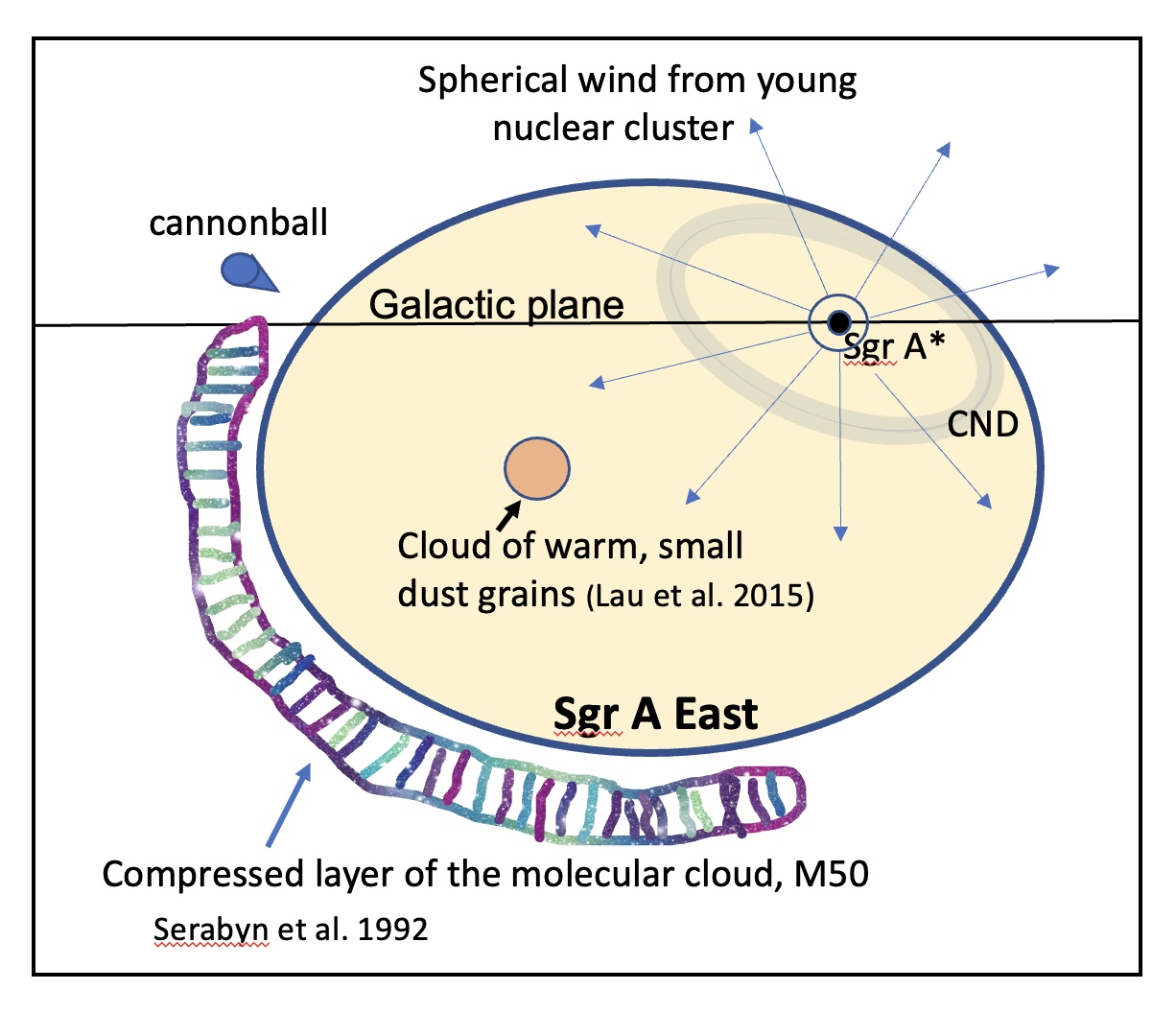}
   \caption{Sgr A East compresses the M50 cloud. \SgrAstar lies at the centre of the young nuclear cluster, which generates a strong wind that is quasi-spherical at large distances. The position of the dust cloud hypothesized to coincide with the site of the explosion is indicated.
     CND stands for the circumnuclear disk.
              }
         \label{sgraeastsqueezingm50}
\end{figure}

The interstellar medium (ISM) at distances smaller than 20 pc from \SgrAstar has several components. These include a circumnuclear disk, which is a dense, clumpy molecular ring having a roughly 
toroidal morphology, and the young nuclear cluster, which is the 
source of a strong nuclear wind. There are also giant molecular clouds, notably the $50\ \mbox{kms}^{-1}$ cloud (M50), which is interacting with Sgr A East, but also the $20\ \mbox{kms}^{-1}$ cloud, which might be in front of Sgr A East and not physically interacting with  it.
The interaction between Sgr A East and the M50 cloud is shown schematically in Fig. \ref{sgraeastsqueezingm50}.
There is also evidence that a  warm, ionised background medium occupies the central volume and rotates with the Galaxy. Other structures in the region are smaller and/or less massive: streamers, ridges (connecting denser clouds), and a mini-spiral constituting the Sgr A West HII region.

\subsection{Gravitational field}

The gravitational field around \SgrAstar, $\Psi (r_{\mathrm{GC}})$, is approximated as spherically symmetric and is generated by the supermassive black hole (SMBH) with a mass of
$4\times 10^6 \MSun$ and by the nuclear star cluster (NSC). The latter consists of two components, both spherically symmetric, NSC1 with a mass $2.7\times10^7 \MSun$, and NSC2 with a mass $2.8 \times 10^9 \MSun$; see \citet{2015MNRAS.447..948C}, \citet{2020A&A...644A..72P}, and \citet{2022MNRAS.510.5266B}. At the considered spatial scales the NSC contribution dominates the gravitational field.
In this gravitational field the rotation velocity $v_{\rm rot}(x,y,z)$ is given as
\begin{equation}
v_{\mathrm{rot}}(x,y,z) = 
\left( \frac{d\Psi (r_{\mathrm{GC}})}{dr_{\mathrm{GC}}}r_{\mathrm{GC}} \right)^{1/2} \, \frac{\sqrt{x^2 + y^2}}{r_{\mathrm{GC}}}
        \label{eq:quadratic}
,\end{equation} 
where $(x,y,z)$ are the galactocentric coordinates, see section \ref{subsec:fitness} below for more details).  The galactocentric distance is  $r_{\mathrm{GC}} = \sqrt{x^2 + y^2 + z^2}$.

\subsection{Model of the ISM}

The proposed simple model of the ISM includes a low-density homogeneous background and two dynamically important components: the M50 molecular cloud and the wind from the young nuclear cluster. All other components (e.g. the circumnuclear disk and the M20 cloud) are omitted.

\subsubsection{Cloud}
\label{subsec:cloud}

The first ISM component is a spherical cloud, representing the M50 cloud. 
It is described by its density profile, size, and the position of its centre relative to \SgrAstar.
Since a large majority of the M50 cloud seems undisturbed by Sgr A East,
our assumptions concern only its southwestern part.

Following \citet{2019MNRAS.484.5734K} and \citet{2022ApJ...929...34M}, we set the 
cloud  density profile as

\begin{equation}
n_{\rm cl}(r) = n_{\mathrm{cl0}} \left(\frac{r}{r_{\mathrm{cl}}}\right)^{-\beta}
\label{m50cloud}
,\end{equation}
where $r$ is the distance from the centre of the cloud, $n_{\mathrm{cl0}}$ is the maximum density $n_{\mathrm{cl0}} = 10^{5.5}\mbox{cm}^{-3}$, $\beta = 5/2$. The mean weight of particles in the molecular medium is taken as $\mu_{\rm mol} = 2.35$. For $r_{\rm cl} = 5\ \mbox{pc}$, 
the cloud has a mass of $7.8\times 10^4\, \mathrm{M}_{\odot}$ at densities above $10^3 \mbox{cm}^{-3}$, and 12.5 \% of the cloud mass has density above $10^5 \mbox{cm}^{-3}$. 

\subsubsection{Nuclear star cluster wind}
\label{subsec:wind}

The second ISM component is the NSC wind.
Its density and velocity profile can be derived in several ways. The adiabatic
approximation of \citet{1985Natur.317...44C} is the natural starting point, but 1D models
can be further improved by adding radiative cooling \citep{2004ApJ...610..226S, 2014ApJ...792..105P}. 3D models \citep{2017ApJ...835...60W} are the next step. In this paper we use the latter and model the wind using the hydrodynamic code FLASH.

The numerical model includes a wind source representing the young nuclear star cluster, mass loading of the star cluster wind from the circumnuclear disk and the fixed gravitational potential of the SMBH, NSC1 and NSC2, as described above. 
The wind is inserted into a sphere with radius $0.5$\,pc, with the mass and thermal energy insertion rates $\dot{M}_\mathrm{YNC} = 1.5\times 10^{-4}$\,$\MSun$\,yr$^{-1}$ and $L_\mathrm{YNC} = 3\times 10^{38}$\,erg\,s$^{-1}$, respectively. This corresponds to $\sim 100$ young massive stars observed in the Galactic young nuclear cluster \citep[e.g.][]{neumayer2020}. Additionally, the wind is mass loaded with rate $\dot{M}_\mathrm{CND} = 4.6\times 10^{-3}$\,$\MSun$\,yr$^{-1}$, that is, the mass-loading factor is $30$. The loaded mass has zero thermal and kinetic energy, and we assume that it originates from the circumnuclear disk \citep[see e.g.][for details of the circumnuclear disk-wind interaction]{blank2016}. The amount of the mass-loaded gas was selected so that the mean wind density approximately corresponds to the density of the background medium in the \SgrAstar vicinity, specifically, the observed radio halo \citep{yzm1987, pedlar1989}. The terminal velocity of the wind resulting from the young NSC wind and the loaded mass is $v_\infty = [2L_\mathrm{NSC}/(\dot{M}_\mathrm{NSC}+\dot{M}_\mathrm{CND})] \simeq 450$\,km\,s$^{-1}$.

The numerical model was implemented within the 3D adaptive mesh refinement (AMR) code FLASH v4.6.2 \citep{fryxell2000}. The hydrodynamic equations were solved using a modified version of the piecewise parabolic method (PPM; Colella \& Woodward 1984) with the time-step controlled by the Courant–Friedrichs–Lewy criterion. The radiative cooling of the gas was calculated using tables provided by \citet{2009A&A...508..751S}, the gas metallicity was assumed to be solar, and the mean particle weight of the ionised medium was set to $\mu_{\rm ion} = 0.609$. Assuming there are enough ionising photons, the gas is not allowed to cool below $10^4$\,K.

\begin{figure*}
   \includegraphics[angle=0,width=0.95\textwidth]{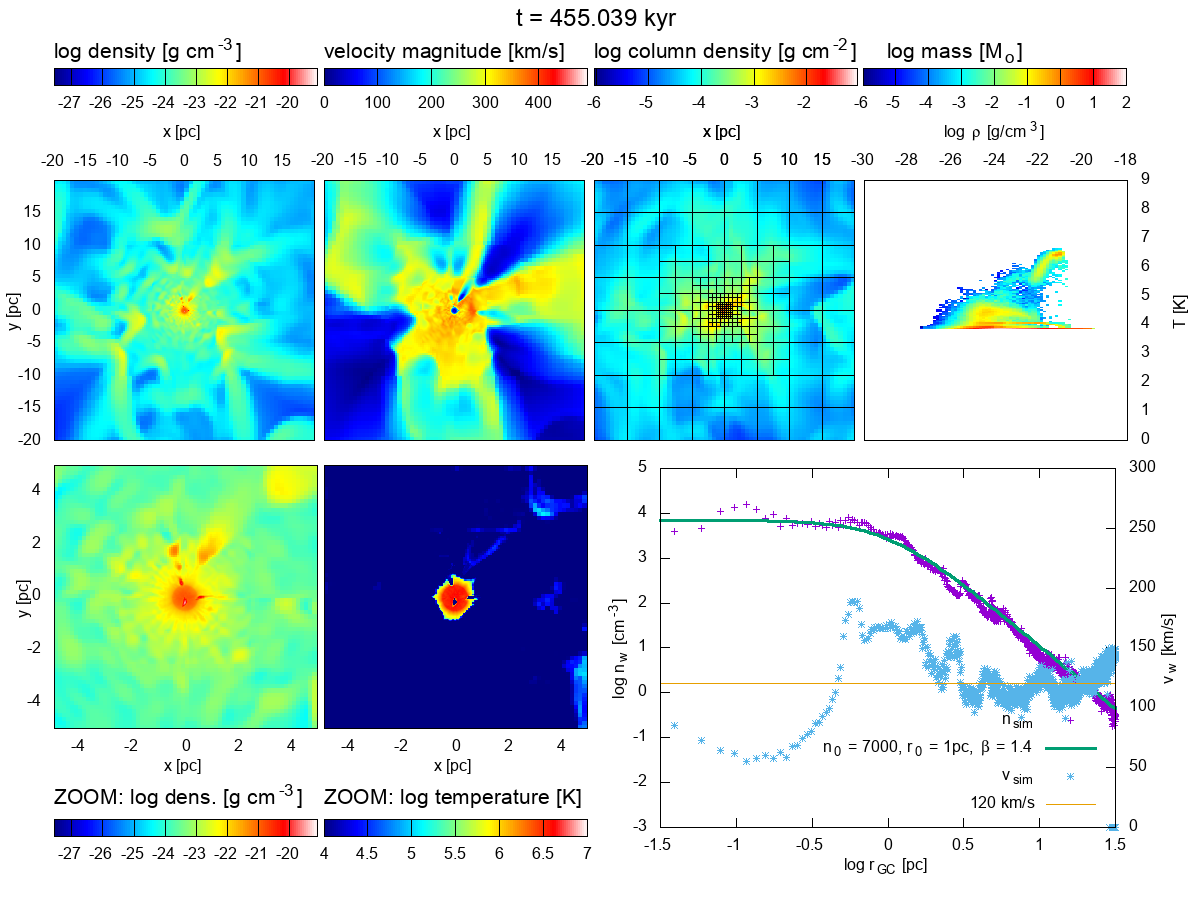}
      \caption{Hydrodynamic simulation of the nuclear star cluster wind at time $455$\,kyr (end of the simulated evolution). The top panels show (from left to right) the logarithm of the gas volume density, the magnitude of the gas velocity, the logarithm of the gas column density, and the gas density-temperature phase diagram, where the colour denotes the gas mass in a given bin. No gas has a temperature below $10^4$\,K since we did not allow the gas to cool below this limit. The volume density and velocity are plotted at plane $z=0$, and the column density is integrated along the $z$-direction. The left two bottom panels show logarithms of the gas volume density and temperature in the central region. The bottom right panel shows radial profiles of the gas density and velocity integrated over axial directions (plus and asterisk, respectively). The solid lines show the fits to these quantities used in the paper.}
         \label{wind}
\end{figure*}

The computational domain was a cube with size $40$\,pc. Initially, it was filled with gas of very low density ($n = 10^{-3}$\,cm$^{-3}$) and high temperature ($T=10^{6}$\,K). An AMR grid with eight refinement levels was used, with the highest resolution in the centre (the smallest grid cells have a size of $0.04$\,pc; see the third top panel from the left in Fig.~\ref{wind} for the grid structure). Technically, the numerical model is very similar to the one described in \citet{2017ApJ...835...60W}.

We let the simulation run until it reached a quasi-stationary state. The wind was thermally unstable due to its high density. This led to (i) a drop in the wind temperature from several million K to $10^4$\,K at radius $\sim 1$\,pc, and (ii) the formation of dense warm clumps that were ejected from the young nuclear cluster. The rarefied warm wind expanded with velocity $300 - 400$\,km\,s$^{-1}$ (slightly below $v_\infty$), but the dense clumps move outwards at lower velocity $\sim 100$\,km\,s$^{-1}$ as they are decelerated by gravity. 

Fig.~\ref{wind} shows the simulation at the end of its evolution $t = 455$\,kyr. The three top left panels show the gas density, velocity, and column density as described in the figure caption. The top right panel shows the distribution of the gas in the density-temperature phase diagram. The two bottom left panels are zooms into the central $5$\,pc showing the gas density and temperature. The bottom right panel shows radial profiles of the mean gas density and the density-weighted mean velocity (both profiles averaged over the two axial directions).

For further calculations with the RING code, we fit the wind mean density with the function
\begin{equation}
n_\mathrm{w}(r_{\mathrm{GC}}) = \frac{n_{\mathrm{w0}}}{\left[1+\left(\frac{r_{\mathrm{GC}}}{r_{\mathrm{w0}}}\right)^2\right]^\beta},
\label{windnsc}
\end{equation}
where $n_\mathrm{w}(r_{\mathrm{GC}})$ is the fitted wind density, and $r_{\mathrm{GC}}$ is the radial coordinate from the Galactic centre. Fitting yields the following parameters: $n_{\mathrm{w0}} = 7000\, \mbox{cm}^{-3}$, $\beta = 1.4$, and $r_{\mathrm{w0}} = 1\ \mbox{pc}$. We approximated the mean wind velocity outside the young nuclear cluster with the constant value $v_{\mathrm{w}}(r_{\mathrm{GC}}) = v_{\mathrm{w0}} = 120$\,km\,s$^{-1}$. The two functions $n_\mathrm{w}(r_{\rm GC})$ and $v_{\rm w0}$ are shown by solid lines in the bottom right panel of Fig.~\ref{wind}.

\subsubsection{Background density}

The low-density background used in our calculations on top of the cloud and the wind is a homogeneous part of the ISM. The default value of 1 cm$^{-3}$ is significantly lower than the density corresponding to the two principal components (molecular cloud and the young nuclear cluster wind), which dominate the tested region, and therefore the background is unimportant. A more thorough discussion of the influence of the homogeneous background is given in Appendix \ref{A2}.

\section{Methods}   

\subsection{RING code}

\begin{figure}
\includegraphics[angle=0,width=0.45\textwidth]{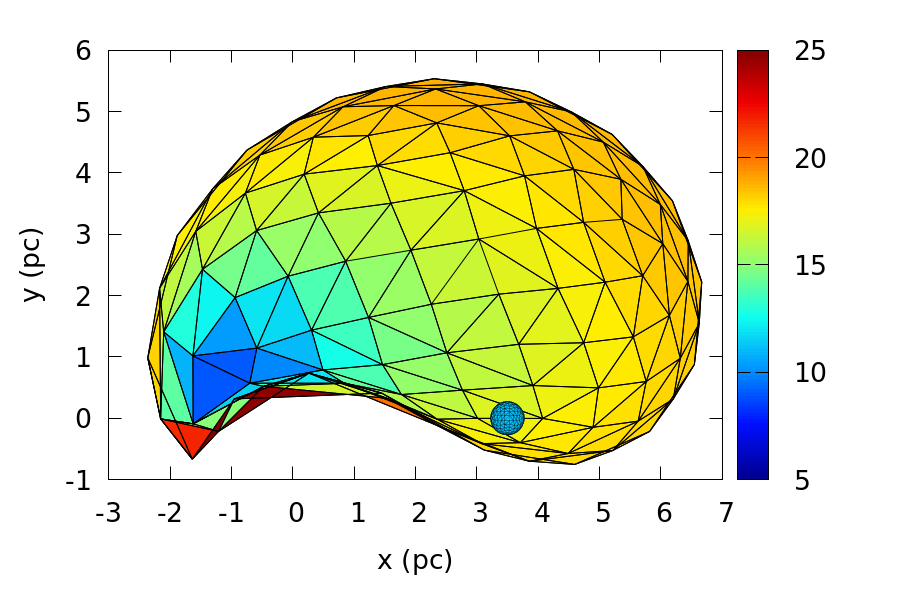}
      \caption{
      RING code, which divides the 2D surface of the 3D expanding structure into $N_{\mathrm{elem}}$ elements. Here, a supernova remnant evolving into a homogeneous medium and stretched by Galactic shear is shown, with the colour depicting the surface density of elements (in $M_{\odot}\mathrm{pc}^{-2}$). The small round blue circle corresponds to the initial position and size of the remnant, and the elongated structure shows the age of 30 kyr.
              }
         \label{elements}
\end{figure}

To reproduce the size, shape, and velocity of expansion of the Sgr A East supernova remnant, we used the hydrodynamic code RING, which is a 3D simplified hydrodynamic code for modelling the evolution and propagation of a compressed shell. The main assumption is the thin-shell approximation, that is, the assumption that the width of the wall of the shell is negligible compared to its diameter, 
and that the shell is dense enough to radiate all thermal energy produced by the shock compressing the ISM. The shell is divided into $N_{\mathrm{elem}}$ elements, which were distributed
along the expanding shell using the HealPix algorithm \citep[details in][]{2020A&A...644A..72P}; see Fig. \ref{elements}. As the thin shell expands, it accumulates mass from the ISM and slows down. RING computes trajectories of shell elements in an external gravitational potential, by solving the equations of momentum conservation:

\begin{equation}
\frac{\mathrm{d}}{\mathrm{d}t}(m_{i} \vec{v_{i}}) + \rho_{\rm out} \cdot [\vec{S_{i}} \cdot (\vec{v_{i}} - \vec{v_{\rm out}})] \cdot \vec{v_{\rm out}} = (P_{\rm int} - P_{\rm out}) \cdot \vec{S_i} + m_i \vec{g}
,\end{equation}

\noindent
where $m_i$, $\vec{S}_i$ and $\vec{v}_i$ are the mass, the surface area vector, and the expansion velocity vector of the ith surface element, respectively.  $P_{\rm int}, P_{\rm out}$ , and $\rho_{\rm out}$ are the internal and external pressures  and the density of the ISM outside of the shell, while $\vec{g} = \nabla \Psi (r_{\rm GC})$ is the local gravitational acceleration. The ISM density 
$\rho_{\rm out} = (n_{\rm bg} + n_{\rm w})\,\mu_{\rm ion} + n_{\rm cl}\,\mu_{\rm mol}$.

Each shell element accretes the encountered ISM if their relative motion is supersonic, 

\begin{equation}
\frac{\rm d}{{\rm d}t}{m_i} = \rho_{\rm out} \cdot [\vec{S}_i \cdot (\vec{v}_i - \vec{v}_{\rm out})]
\label{eq:accr}
.\end{equation}

\noindent
The pressure behind the shell is described as

\begin{equation}
P_{\rm int} = \frac{2 E_{\rm th}}{3 V_{\rm int}}.
\end{equation}

\noindent
By disregarding  radiative cooling within the cavity behind the shell, the change of $E_{\rm th}$ is given by the adiabatic energy balance equation,

\begin{equation}
\frac{{\rm d} E_{\rm th}}{{\rm d} t} = - \frac{{\rm d} V_{\rm int}}{{\rm d} t} \cdot P_{\rm int}
\label{eq:enercav}
,\end{equation}

\noindent
where $P_{\rm int}$ and $V_{\rm int}$ are the inner pressure and the volume of the cavity.

The equations above provide a much simpler description than fully hydrodynamic codes. The advantage of the RING code is that it is much faster than other hydrodynamic codes, which facilitates exploration of a multi-dimensional parameter space. 

The theoretical background of this method was developed by 
\citet{1995RvMP...67..661B}; it was used to match observational data by 
\citet{2018A&A...619A.101E} and others. A more detailed description of the code RING is given by \citet{2020A&A...644A..72P}.

\subsection{Fitness function}
\label{subsec:fitness}

To estimate the correctness of the numerical model, we defined a fitness function that describes
how well the model resembles the observed properties of Sgr A East. For this reason, we needed
a precise but simple characterization of the shape, orientation, and size of the Sgr A East supernova remnant. No other constraints such as on the age or mass of the remnant were enforced in the model. 

Based on Figs. \ref{SgrAE} and \ref{sgraeastsqueezingm50}, we demanded that the major
axis of the structure be parallel to the Galactic plane. The extent of Sgr A East can be defined either using the elliptical fit mentioned in the Introduction, or simply by using the total extent of the remnant in the $x-$ and $z$ -coordinates defined in the next paragraph. We used the latter approach because it allows for more distorted shapes of the remnant and is easier to implement.

We defined a Cartesian coordinate system in which the $x$-coordinate was parallel to the Galactic equator and the $z$-coordinate was perpendicular to the Galactic plane; its centre was \SgrAstar. Angular coordinates of the rectangle around Sgr A East in this system  are $(-2.4'; -2.0')$ for a lower left corner and $(1.5'; 0.8')$ for an upper right corner. The orientation is the same as in Fig. \ref{sgraeastsqueezingm50}: the M50 cloud lies below the Galactic plane and to the left from \SgrAstar. 
The selected area is a rectangle
that is about 10 \% larger than the dimensions of the ellipse $3.5'\times2.5'$, which are taken as canonical values for Sgr A East.

For $R_{\odot} = 8.2\ \mbox{kpc}$, the minimum and maximum values of the rectangle drawn around Sgr A East are
$x_{\rm min}^{\rm obs}= -5.8\ \mbox{pc}$, 
$x_{\rm max}^{\rm obs}= 3.3\ \mbox{pc}$, 
$z_{\rm min}^{\rm obs}= -4.8\ \mbox{pc}$, and 
$z_{\rm max}^{\rm obs}= 1.8\ \mbox{pc}$. 
The $y$-coordinate in this system is parallel to the line of sight and increases with increasing distance. The ratio of the $x$ and $z$ size is 1.4.

For each calculated model at each time, the fitness function evaluated 
\begin{enumerate}
    \item \textbf{the size and the position:} differences between the desired values of $x_{\rm min}^{\rm obs}$, $x_{\rm max}^{\rm obs}$, $z_{\rm min}^{\rm obs}$ , and $z_{\rm max}^{\rm obs}$ and the values computed from the model; 
\item \textbf{the orientation:} the inclination $\phi$ of the major axis of the computed structure to the Galactic equator;  
\item \textbf{the shape:} the comparison of the computed $\frac{\Delta x}{\Delta z}$ to the observed ratio between the $x$ and $z$ size; 
$\Delta x = x_{\rm max} - x_{\rm min}$, and $\Delta z = z_{\rm max} - z_{\rm min}$. 
\end{enumerate}

Each component of the fitness function was normalized and given some weight, and then the components were added together (see Appendix \ref{A1}). We experimented with weights of individual components of the fitness function, but results were not dependent on the exact choice of weights and were quite robust.
 
\section{Results}
\label{sec:results}

This section describes the grid of calculations performed with the code RING, with the aim of reproducing the morphology of the supernova remnant Sgr A East. Simulations were made without the external gravitational field. This choice increases the speed of calculations; the effect it has on simulations is discussed in Sect. \ref{sec:discussion}.

The ISM has two components: the molecular cloud, and the NSC wind. Their density and velocity distributions are given in Sects. \ref{subsec:cloud} and \ref{subsec:wind}; a low-density background ($n_{\mathrm{bg}} = 1\ \mathrm{cm^{-3}}$) was added to avoid zero-density regions.

There are seven free parameters: the 3D position of the molecular cloud ($x_{\rm cl}$, $y_{\rm cl}$, $z_{\rm cl}$) and its radius $r_{\rm cl}$, which, together with Eq. \ref{m50cloud}, also gives its mass; and the position of the supernova explosion ($x_0$, $y_0$, $z_0$). We constrained the range of possibilities for the position of the SN explosion and the cloud from observations. 

Sgr A East has a relatively regular shape and is probably young. The explosion therefore took place not far away from the current centre of Sgr A East, and certainly inside the remnant.  To be on the safe side, we allowed the explosion to have taken place outside the current Sgr A East outline, but not very far, that is, $x_0$ and $z_0$ must lie in the interval (-12,0) pc (see Fig. \ref{best-xclzcl-x0z0}). The third coordinate ($y_0$) is not known, but we placed the explosion behind \SgrAstar to be consistent with observations, up to a distance of 12 pc, that is, $y_0$ must lie in the interval (0,12) pc.

The position of the cloud was  varied in similar intervals, but we only allowed configurations in the $x-z$ plane where the explosion is closer to \SgrAstar than the cloud centre (this condition does not apply to the $y$ -coordinate). There are detailed observations of the M50 cloud extending well beyond our computational domain, but since we are only interested in its southwestern part, in which the interaction between the cloud and the supernova remnant takes place,
they are not so helpful in prescribing the position of our spherical cloud. The radius of the cloud was varied between 5 and 8 pc.

Other parameters needed for our simulations, such as the energy of the explosion, were held
constant ($E_{\mathrm{SN}} = 10^{51}\ \mbox{erg}$). Each calculation was followed till 30 kyr. At each kyr, the position and shape of the evolving supernova remnant (more precisely, the
maximum and minimum values of its $x-$, $y-,$ and $z$ -coordinates and the angle between its major axis and the Galactic equator) and its mass were recorded. These data were then used for the fitness calculations.

\subsection{Three types of ISM scenarios}

As described above, our ISM has two components: wind, and cloud. A low-density background ($n_{\rm bg} = 1\, \mbox{cm}^{-3}$) was added mostly for numerical reasons, it does not influence the shape of the supernova remnant. To check the relevance 
of the two important components, we also performed calculations in an ISM that contained only one of these components. These three scenarios are called wind+cloud
(where both components are present), cloud-only (with only a molecular cloud, without
the NSC wind), and wind-only (with only the wind, but without a molecular cloud).

The simulations in the wind+cloud and cloud-only scenarios were made in the same grid
of seven parameters described above. The simulations for the wind-only scenario  were performed
in a restricted grid because the parameters of the cloud are not relevant in this case.

\begin{figure}
   \includegraphics[angle=0,width=0.45\textwidth]{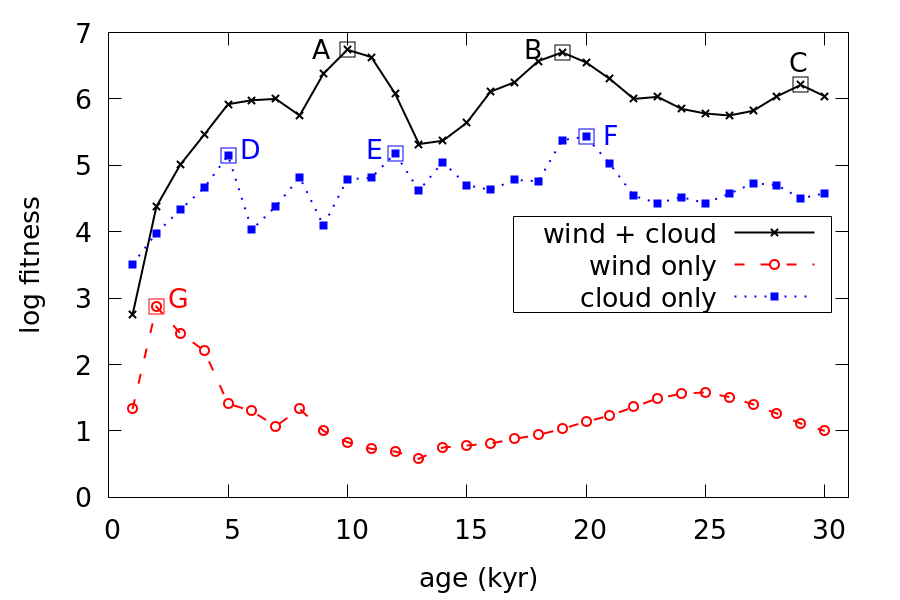}
      \caption{Logarithm of the fitness of the best model at a given age (x-axis)
      for the wind+cloud models (black), cloud-only models (blue), and wind-only models (red).
      Specially designated models (A-G) refer to models discussed in the text
      and in other figures.
              }
         \label{fitness_best_at_age}
\end{figure}

\subsection{Locations of the explosion and the cloud}

Figure \ref{fitness_best_at_age} gives the fitness for the best model as a function of age. 
The best model was selected as the simulation at a given age that fits best. Therefore, the parameters of the best model (position of an SN explosion, position
and size of the cloud) are not constant with time. Fig. \ref{fitness_best_at_age}
clearly shows that the best results are reached for the wind+cloud ISM scenario.
On the other hand, simulations containing only the wind (wind-only scenario) 
fit only poorly; we were unable to create both the observed elongation and 
the position of the Sgr A East remnant with this model. Cloud-only models are better, but still worse than the combined wind+cloud models. The fitness of the best model is not a monotonic function of age; all curves in Fig. \ref{fitness_best_at_age} have local maxima. These maxima are designated models A-G and are discussed below.

The best model (A) for the wind+cloud scenario has an age of 10 kyr, but there are
also quite good solutions at older ages (19 kyr for B and 29 kyr for C). Based only on results of our simulations, we cannot rule out ages for Sgr A East older than 10 kyr.
On the other hand, ages younger than $\sim $ 5 kyr do not provide satisfactory solutions.

\begin{figure}
   \includegraphics[angle=0,width=0.45\textwidth]{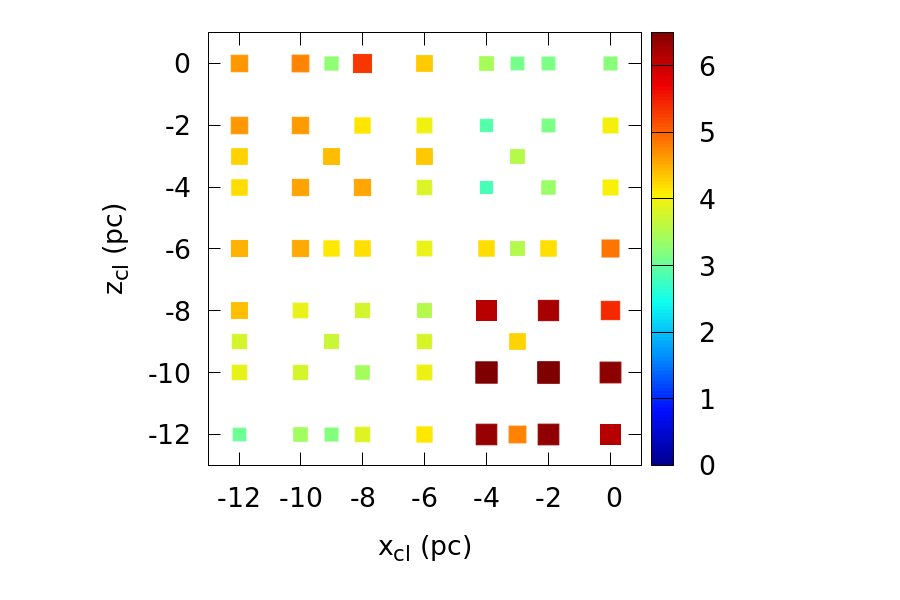}
   \includegraphics[angle=0,width=0.45\textwidth]{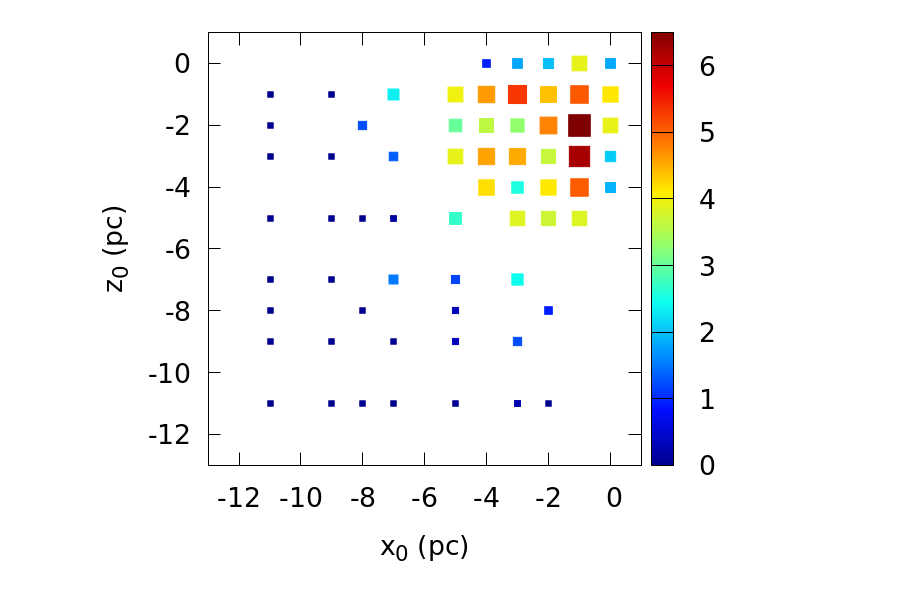}
      \caption{Wind+cloud scenario: Best models for a given $(x_{\rm cl},z_{\rm cl})$ 
      position of the M50 cloud (top)
      and best models for a given $(x_{0},z_{0})$ position of the supernova explosion
      (bottom). The colour and the size of squares both reflect the fitness of the best
      model at the indicated coordinates (the larger and the redder, the better).
              }
         \label{best-xclzcl-x0z0}
\end{figure}

Figure \ref{best-xclzcl-x0z0} shows best models for a given position $(x_{\rm cl},z_{\rm cl})$ 
of the M50 cloud and for a given position $(x_{0},z_{0})$ of the supernova explosion.
It demonstrates that good models 1) prefer the cloud M50 to be located relatively
close to the Galactic centre in the $x$ coordinate and farther away and below it ($z$), and 2) have quite a tightly constrained position of the SN explosion (there are no good models outside a small region in the $xz$ plane).

\begin{figure*}
   \includegraphics[angle=0,width=0.3\textwidth]{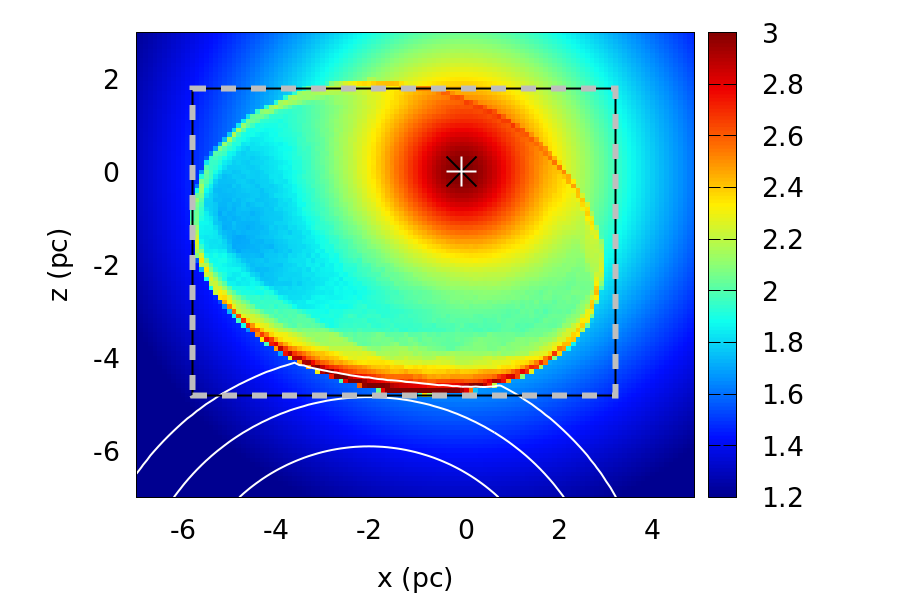}
   \includegraphics[angle=0,width=0.3\textwidth]{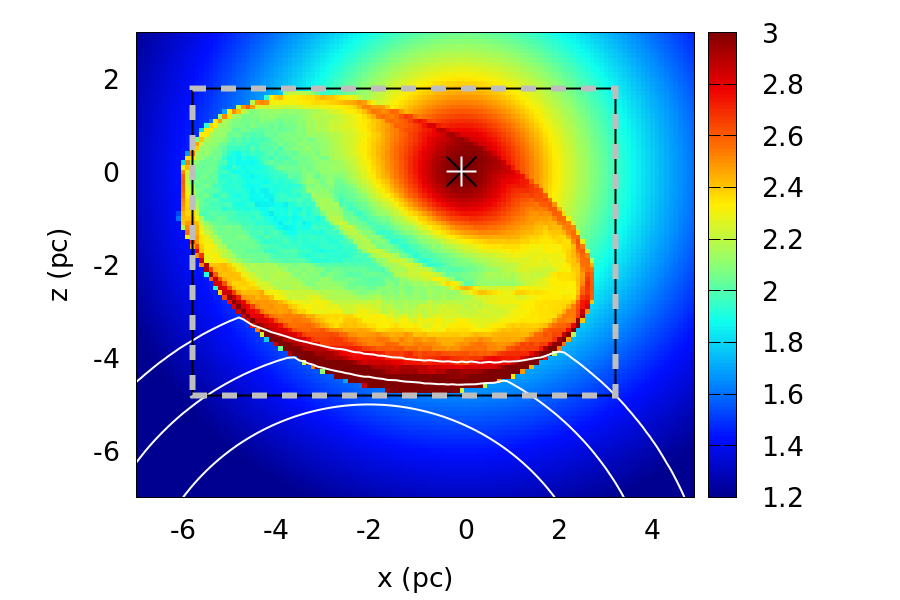}
   \includegraphics[angle=0,width=0.3\textwidth]{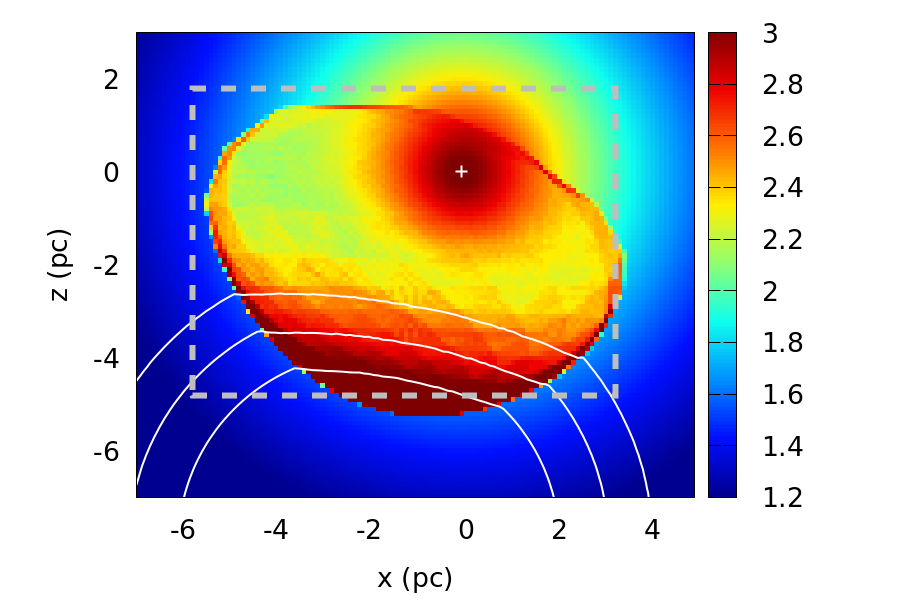}
   
   \includegraphics[angle=0,width=0.3\textwidth]{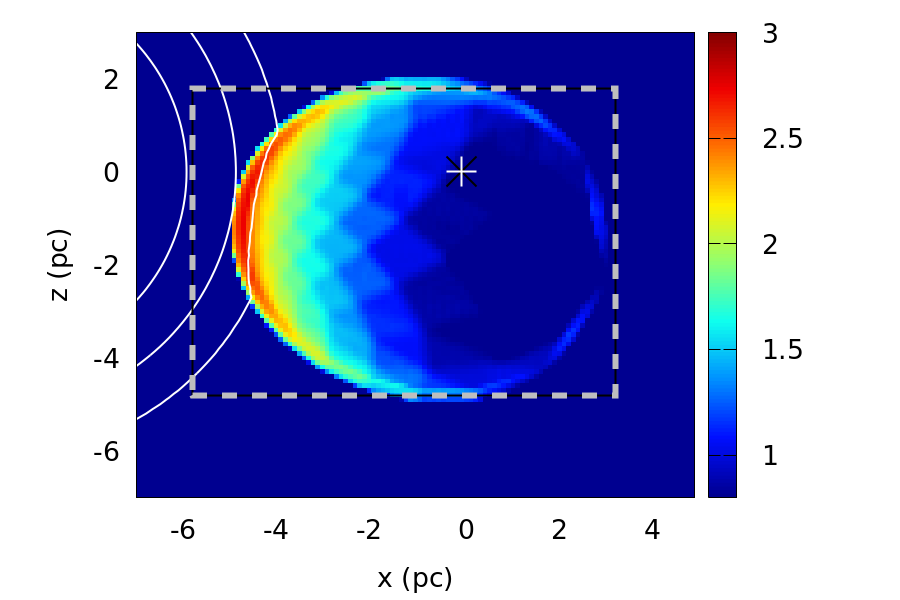}
   \includegraphics[angle=0,width=0.3\textwidth]{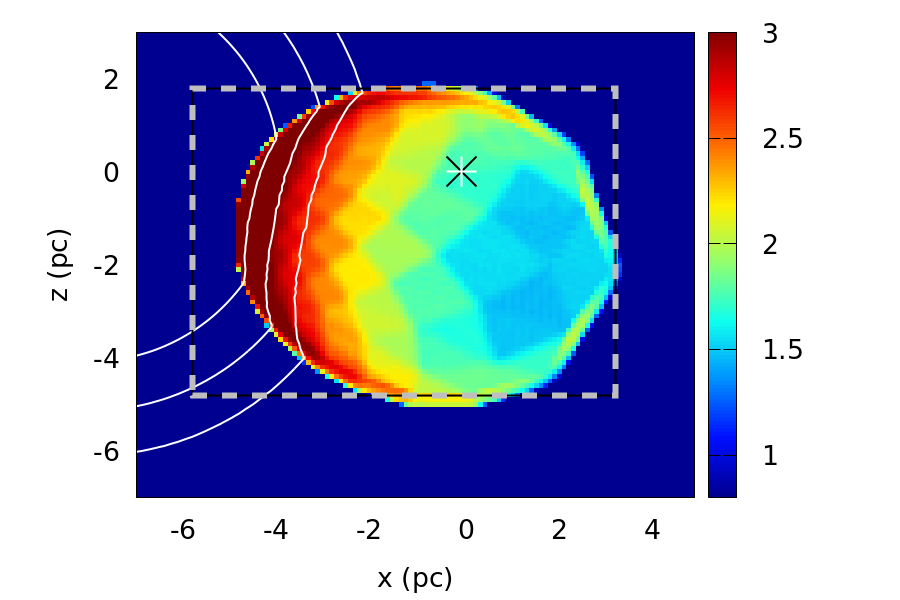}
   \includegraphics[angle=0,width=0.3\textwidth]{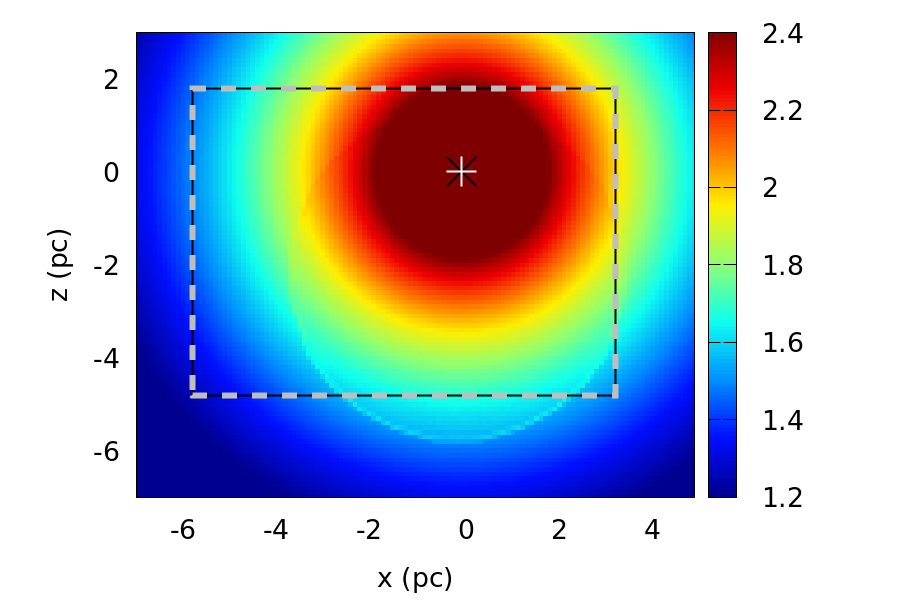}   
   \caption{Best models of Sgr A East found by our method, projections in the $xz-$plane.
     First row, from left to right: Wind+cloud scenario, models A (10 kyr), B (19 kyr),
      and C (29 kyr). Second row, from left to right: Cloud-only scenario, 
      models D (5 kyr) and F (20 kyr); wind-only scenario G (2 kyr). The colour shows
      the (logarithm of the) column density of the warm ionised medium (in arbitrary
      units), and white contours show the column density of the molecular medium 
      (logarithmic scale in arbitrary units). The dashed rectangle denotes the
      observed position of Sgr A East, and the black and white cross shows \SgrAstar at the
      centre of the Galaxy.
              }
         \label{model_3pan}
\end{figure*}

Figure \ref{model_3pan} shows six (out of seven indicated) good models from Fig. \ref{fitness_best_at_age}.
The upper row shows models A, B, and C in the wind+cloud scenario, and the lower row shows
models D and F in the cloud-only scenario, and model G for the wind-only scenario. 
The colour indicates the column density of the warm ionised medium, and the contours show the 
column density of the molecular medium (= M50 cloud). The parameters of the models are given
in Table \ref{models-1}.

The wind+cloud scenario places the supernova explosion at a position below and to the left
from \SgrAstar (around 2 pc below the Galactic plane and 1 pc to the left from the centre), 
and behind it (around 3 pc behind \SgrAstar).
It favours the position of the M50 cloud slightly to the left of \SgrAstar (the best model
has $x_{\rm cl} = -2$ pc), below the plane (-10 pc), and slightly behind it (2 pc). 
Fig. \ref{model_3pan} shows that the interaction zones between the cloud and SN expanding shell as well as between the wind and SN expanding shell are brighter than the rest of the remnant.

The cloud-only scenario places the supernova explosion more to the left of \SgrAstar and 
closer to the Galactic plane than the previous case  (-3 pc in the $x-$coordinate, -1 pc in $z$), 
and the explosion is located at the same distance as that of \SgrAstar 
(this is a difference to the wind+cloud scenario). The cloud M50
must be located only to the left of \SgrAstar. This scenario creates structures
with one very pronounced (= bright) hemisphere; see Fig. \ref{model_3pan}.

The M50 cloud may lie somewhere between the positions found by best
models in the wind+cloud and cloud-only scenarios. Because of this, and because the real
cloud is not spherical and we only trace its one rim in our simulations, we cannot use it to prioritize one of our two scenarios.
However, observations show that Sgr A East and its centre of explosion 
 lie behind Sgr A* \citep{1987AJ.....94.1178Y,1989ApJ...342..769P,1999ApJ...512..230Y}, which favours our wind+cloud scenario.

\begin{table}
\begin{tabular}{lrrrrrrrr}
\hline
model & $x_{\rm cl}$ & $y_{\rm cl}$ & $z_{\rm cl}$ & $r_{\rm cl}$ & $x_0$ & $y_0$ & $z_0$ & age \\
~     & pc & pc & pc & pc & pc & pc & pc & kyr \\
\hline
A & -2 & +2 & -10 & 5 & -1 & +3 & -2 & 10 \\
B & -2 & 0 & -10 & 6 & -1 & +1 & -2 & 19 \\
C & -2 & +4 & -8 & 5 & -1 & +2 & -3 & 29\\
\hline
D & -10 & 0 & 0 & 5 & -3 & 0 & -1 & 5 \\
E & -10 & 0 & 0 & 6 & -3 & +2 & -1 & 12 \\
F & -8 & 0 & 0 & 5 & -3 & 0 & -1 & 20 \\
\hline
G &  &  &  &  & 0 & +10 & -2 & 2 \\
\hline
\end{tabular}
\caption{Parameters of the best models of the wind+cloud scenario (A, B, and C), cloud-only scenario (D, E, and F), and the wind-only scenario (G).}
\label{models-1}
\end{table}

\begin{figure*}
   \includegraphics[angle=0,width=0.3\textwidth]{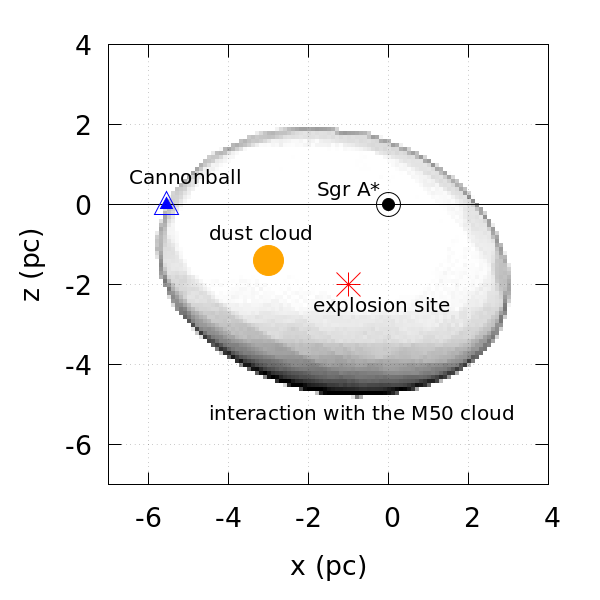}
   \includegraphics[angle=0,width=0.3\textwidth]{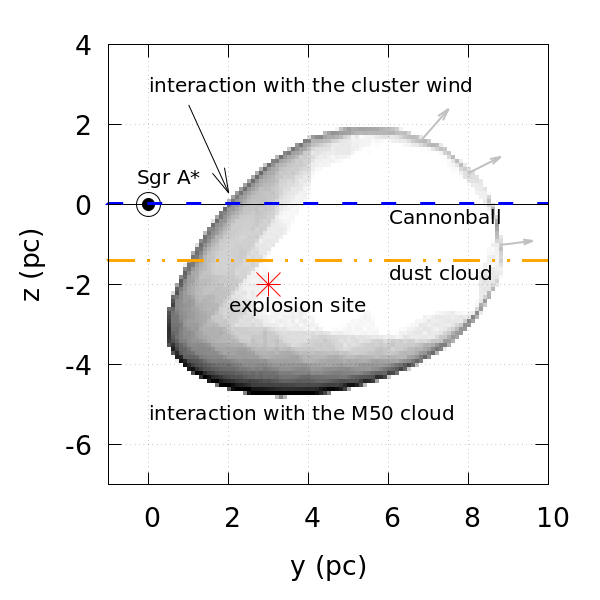}
   \includegraphics[angle=0,width=0.3\textwidth]{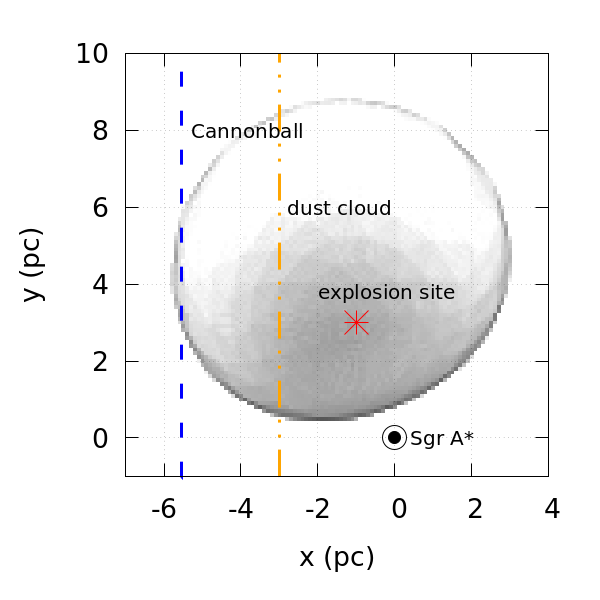}
      \caption{Projections of the best model A in $xz$-plane (left panel; the view from the Sun), in the $yz$-plane (middle panel), and in the $xy$-plane (right panel; view from the Galactic pole), together with the position of other objects: \SgrAstar with the young nuclear cluster wind around it (a double black circle), the Cannonball pulsar (the double blue triangle or the dashed blue line), and the dust cloud from \citet{lau2015}, the orange circle or the dot-dashed orange line.
      The sizes of symbols are not to scale.
      Small grey arrows indicate an expansion into the lower-density ISM.
      Although displaced, the box size is the same in all panels,
      and \SgrAstar appears at position 0,0 in all panels.
              }
         \label{model_xyz}
\end{figure*}

\begin{figure}
   \includegraphics[angle=0,width=0.45\textwidth]{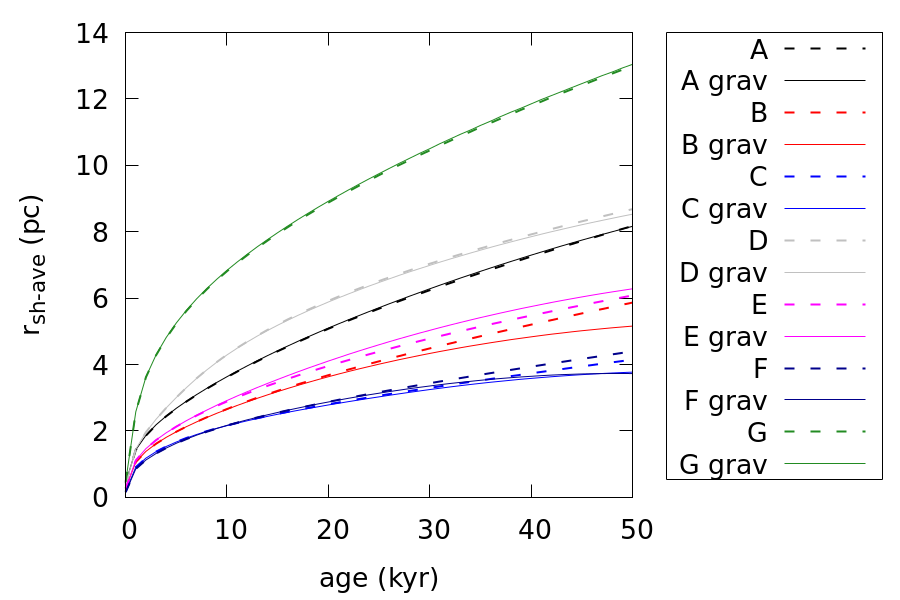}
      \caption{Comparison between average sizes of shells calculated with (solid lines) and without (dashed lines) the gravitational field. Models A, B, C, D, E, F, and G are shown.
              }
         \label{nograv-vs-grav}
\end{figure}

Fig. \ref{model_xyz} shows the calculated shape of the supernova remnant (our best model A) together with positions of the supernova explosion; the dust cloud from \citet{lau2015}, which is  a previously suggested location of the supernova explosion as mentioned in the Introduction; \SgrAstar ; and the Cannonball pulsar. All three projections, $xz$, $xy,$ and $yz$, are shown.

In the $xz$-projection, corresponding to our view from the Sun, the supernova explosion site does not coincide with the position of the dust cloud.
However, there is an interesting alignment of the centre of the explosion, the dust cloud, and the Cannonball pulsar. If the dust indeed originates in the supernova explosion, it could inherit a part of the initial inertia of the progenitor star. A (projected) velocity of about 200 $\mathrm{kms}^{-1}$ would be needed to travel from the explosion site to the observed position of the dust cloud. Alternatively, the dust could gain momentum out of the explosion.

The $yz$-projection shows that the supernova remnant is squeezed on one side by the molecular cloud and on the other by the young nuclear cluster wind. It is expanding away from these two ISM components and from \SgrAstar. The view from the Sun does not indicate much about this shape.

\section{Discussion}
\label{sec:discussion}

This section describes the impact of the simplifications used in our simulations. We discuss the omission of the gravitational field, the existence of other ISM components, and our treatment of the extent of Sgr A East. The section also gives results for the supernova remnant expansion into a homogeneous ISM taken from previous papers, and analyses the mass infall to the Galactic centre caused by our model of Sgr A East.

\subsection{Gravitational field}
\label{subsec:grav}

\begin{figure}
   \includegraphics[angle=0,width=0.45\textwidth]{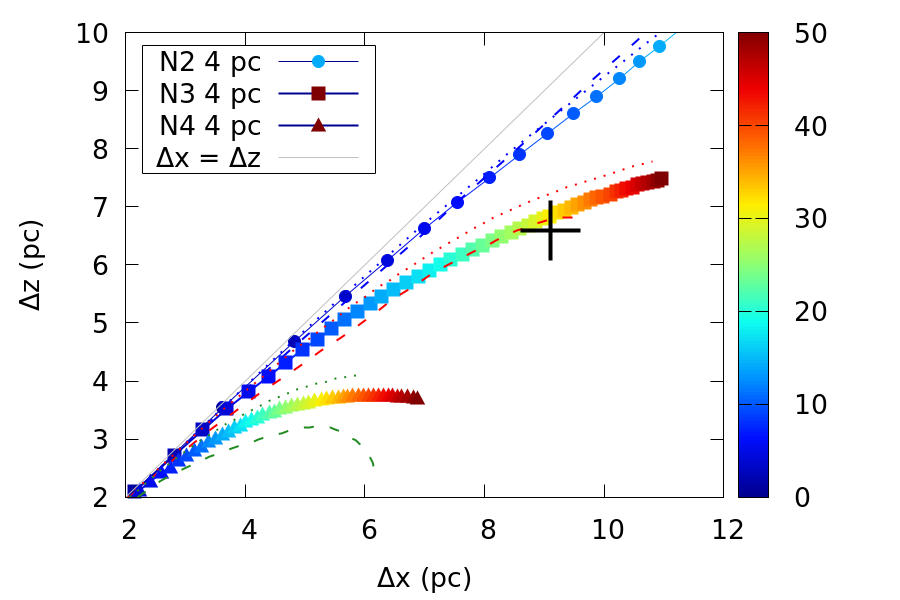}
      \caption{Evolution of the dimensions ($\Delta x$, $\Delta z$) of the supernova remnant evolving into 
      a homogeneous medium with densities $10^2$, $10^3$ , and $10^4$\ cm$^{-3}$ (models N2, N3, and N4) in the gravitational field described in Sect. 2.
      Coloured symbols are for a galactocentric distance $r_{\rm GC} = 4\ \mbox{pc}$, and lines
      belong to distances of $r_{\rm GC} = 2\ \mbox{pc}$ (dashed) 
      and $r_{GC} = 6\ \mbox{pc}$ (dotted) for
      each density (circles for N2, squares for N3, and triangles for N4). Colour denotes the remnant age (0-50 kyr).
      The grey line is $\Delta x = \Delta z$, and the black cross shows the observed dimensions of Sgr A East.
              }
         \label{hmg_grav_novelo}
\end{figure}

The importance of the gravitational field is explored in Fig. \ref{nograv-vs-grav}, in which we compare the average size of a shell as a function of time for all our selected models. The differences between models with and without an external gravitational field are very small, in particular for ages younger than 30 kyr. All results quoted in Sect. \ref{sec:results} are based on simulations with ages $\leq 30$ kyr. Therefore, our estimates of the positions of the M50 cloud and the explosion site of the SN are not affected by the omission of the external gravitational field.

\subsection{Homogeneous medium}
\label{subsec:hmg}

Previous works \citep{1998IAUS..184..317U, maeda2002} have tried to explain the shape of Sgr A East by the shear in the velocity field near the Galactic centre. We have shown in the previous paragraph that the gravitational field is not very important for the (early) evolution and shape of our models. However, the ISM density gradients in the wind+cloud scenario used in Fig. \ref{nograv-vs-grav} are relatively large and dominate especially the early evolution of the remnant.  The situation for a supernova remnant expanding into a homogeneous medium might be different. There, the shape is determined by the gravitational field, which 1) stretches the structure in the Galactic plane by  differential rotation, and 2) flattens it in the $z$-direction by the $K_z$ force perpendicular to the plane. In theory, these two forces both have the desired effect because Sgr A East has a ratio of its dimensions of ${\Delta x}/{\Delta z} \simeq 1.4$ (see Sect. \ref{subsec:fitness}). Therefore, it is natural to ask whether it is possible to account for the observed morphology of Sgr A East by assuming that it is expanding into a homogeneous medium.

However, gravitational forces need some time to have an effect. Fig. \ref{hmg_grav_novelo}
shows the $x$- and $z$-extent ($\Delta x$, $\Delta z)$ of a remnant evolving into a homogeneous density medium ($10^2$, $10^3$ , and $10^4$ cm$^{-3}$) at the initial distance of 4 pc from the centre (different symbols for different densities), with the evolution 
at initial distances of 2 and 6 pc indicated by dashed lines. The $x$-extent depends on the viewing angle and the starting position (azimuth), so that the figure shows the simulation with the highest possible value of $\frac{\Delta x}{\Delta z}$. The black cross in Fig. \ref{hmg_grav_novelo} shows the sky-plane dimensions of Sgr A East.

 Fig. \ref{hmg_grav_novelo} clearly shows that structures evolving in the $10^2$ cm$^{-3}$ environment are too spherical or too large to resemble Sgr A East. In contrast, experiments at a density of $10^4$ cm$^{-3}$ lead to much smaller and more elongated structures. This is valid for all distances, $r_{\rm GC}$, of the supernova explosion below 7 pc.

The case with a density of $10^3$ cm$^{-3}$ seems to be the most promising for creating a properly elongated remnant with the right size. Unfortunately, the time required to achieve the observed size is around 30 kyr, which is longer than the expected age of Sgr A East, which is 10 kyr at most and may be younger.

Previously, \citet{Uchida+98} reported 2D simulations of Sgr A East using a version of the thin-shell approximation. They explained the flattening of Sgr A East by the shear associated with Galactic rotation. They did not estimate the age of the remnant, but reported that
the influence of the shear should be visible on timescales of a few tens of kiloyears.
We broadly agree with this statement,
but Sgr A East is probably younger than several tens of thousands of years. The shear therefore cannot be the sole reason behind the observed elongation of the remnant.

\subsection{Initial momentum}

\begin{figure}
   \includegraphics[angle=0,width=0.45\textwidth]{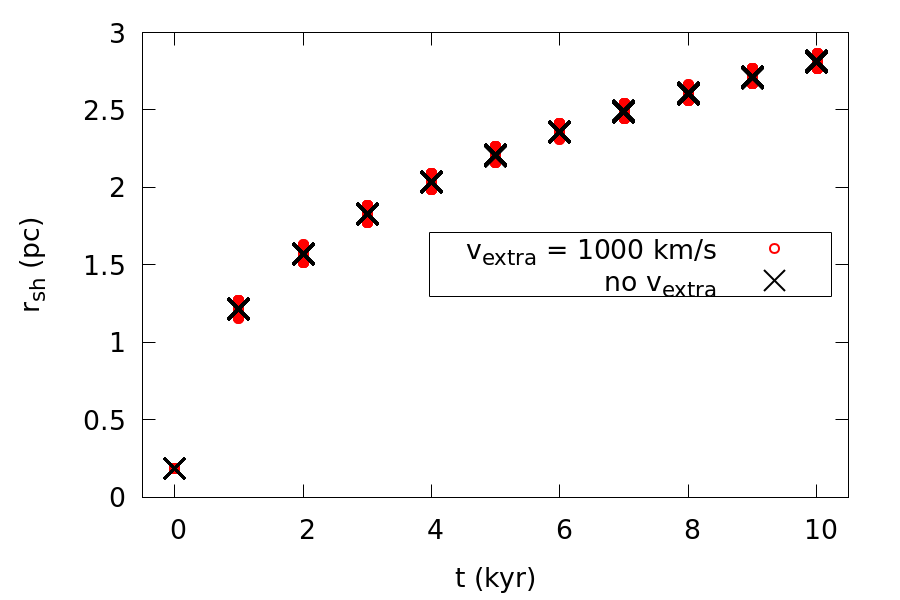}
      \caption{Comparison of the distance of shell elements from the centre of
      the supernova remnant for the homogeneous case without an added velocity
      (black crosses) and the case with an additional velocity of the ejecta 
      (in this case, $1000\ \mbox{kms}^{-1}$ was added in the $x$-direction; red circles).
              }
         \label{inimom_n3x3}
\end{figure}

When we give the supernova ejecta some additional motion (inherited from the additional motion
of its predecessor), we can create elongated structures. However, unless the initial velocity is unrealistically high, this added momentum is low compared to the resulting
momentum of the supernova remnant. This is because the initial mass of the
ejecta is low compared to the mass accumulated from the ambient medium. Therefore, we cannot expect large  effects from the additional momentum carried by the supernova progenitor.

This is indeed the case. Fig. \ref{inimom_n3x3} shows a comparison of two cases,
one with the additional velocity, and one without it. A velocity of 
$1000\ \mbox{kms}^{-1}$ is added in the $x$-direction, which leads to a slight non-sphericity of the remnant, but this effect is very small. After all, the additional kinetic energy in the case with the additional velocity was only $10^{49}\ \mbox{erg}$, which is far lower
than the supernova energy. Similar results were reached when the initial momentum
was added in any other direction.

To summarize, we cannot expect to explain the high ellipticity of Sgr A East by the additional motion of the SN predecessor alone. The value of 1000 $\mbox{kms}^{-1}$ shown above is far
higher than the typical velocity of a star in the nuclear star cluster (expected values are around 100 $\mbox{kms}^{-1}$), therefore, no significant effect should be expected.

\subsection{Other ISM components}

Our ISM model includes the wind from the nuclear star cluster and the giant molecular cloud M50.
There are other components in the ISM around \SgrAstar, such as gas streams, the circumnuclear disk, and other molecular clouds (see Sect. \ref{sec:inside20pc}). 

The circumnuclear disk, one of the most obvious objects found close to \SgrAstar,
lies inside our young nuclear cluster wind region, but subtends only a small solid angle and therefore does not block the wind. From this perspective, we can ignore the circumnuclear disk since the expansion of Sgr A East in the direction toward \SgrAstar is already impeded by the wind and the circumnuclear disk would not bring new features to our simulations. Others have come to a similar conclusion (e.g. \citet{rockefeller2005}).

A potential exception to our assumption that other ISM components have little impact on the shape of Sgr A East could be the presence of another molecular cloud, for instance the M20 cloud. It is not altogether clear whether this cloud interacts with Sgr A East.
It is often expected to lie in a different region than the M50 cloud and Sgr A East,
probably in front of \SgrAstar; see, for example \citet{2000ApJ...533..245C}.
As we are interested in the simplest scenario, we assumed that the M20 cloud is at least currently 
outside the interaction zone, and therefore, we did not take it into account.

\subsection{Slightly smaller Sgr A East}

We searched our best models on the basis of the total extent of the
6cm VLA observations because we wished to include all emission from the remnant 
(see Sect. \ref{subsec:fitness} for an explanation). However, we may also try to reduce
the size of the rectangle encompassing Sgr A East, so that it encloses just the ellipse delineated 
in Fig. \ref{SgrAE}. This ellipse is at least partially defined by the positions of (local)
maxima in an observed brightness map, which might be connected to the walls of the remnants
with the highest projected densities. After adjusting $x_{\rm min}$, $x_{\rm max}$,
$z_{\rm min}$, and $z_{\rm max}$ to these lower values, we searched for the best models
in our computational grid. All models A-F selected on the basis of the
larger extent were also selected when these lower values were used, but their ages were
younger by 1 kyr. Relative fitness values among models (as shown in
Fig. \ref{fitness_best_at_age}) are slightly different, but all the best models were selected
again and no new ones emerged.

\subsection{Mass delivery to the central 1 pc}

\begin{figure}
   \includegraphics[angle=0,width=0.45\textwidth]{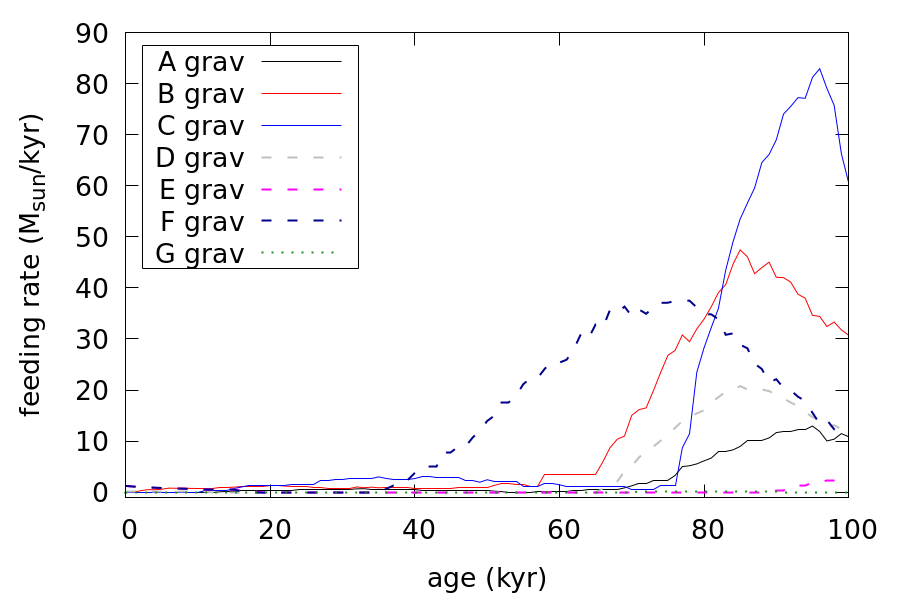}
      \caption{Feeding rates (the rate at which the mass falls to the central 1 pc)
      for models A, B, C, D, E, F, and G.
              }
         \label{feed-rate}
\end{figure}

We evaluated the feeding rates, that is, how much mass is pushed to the inner 1 pc of the Galaxy, and the times at which the mass is pushed,
for selected models (see Fig. \ref{feed-rate}). For all models (A-F), the feeding rates are very low before the desired size, spatial orientation, and shape were reached, that is,
lower than $10^{-3}\MSun \mathrm{yr}^{-1}$. In the subsequent evolution, however, the feeding increased by one or two orders of magnitude.  In the case of models A, B, and C (wind+cloud scenario), the rates are about $10^{-2}-10^{-1}\MSun \mathrm{yr}^{-1}$ at the age of 70-80 kyr. This means that the mass delivery to the vicinity of the Galactic centre caused by the supernova Sgr A East should increase in about 70 kyr from now.    

The activity of the Galactic centre is connected to the mass infall to \SgrAstar. This activity
is currently at a very low level, but it changes with time \citep[see e.g.][]{2014IAUS..303..333P}. 
The increased mass infall from Sgr A East to the circumnuclear disk, which we assume will happen in about 70 kyr, will presumably increase the overall activity of our Galactic centre. In total, the supernova remnant will bring about $1\,000\,\MSun$ of matter into the inner parsec around \SgrAstar. A part of it should join the accretion flow toward the SMBH, depending on the processes taking place in the circumnuclear disk, and will have observable consequences.

\section{Conclusions}

We have modelled the size, shape, and position of the supernova remnant Sgr A East
in the vicinity of the Galactic centre using simplified hydrodynamical simulations. 
The properties of Sgr A East were derived from radio observations at 6 cm wavelength (VLA).
The ISM was modeled with two components: a molecular cloud resembling the observed M50 molecular cloud, and the wind from the young nuclear star cluster. Based on previous calculations, we also  analysed simulations of a supernova remnant in a homogeneous ISM.

\begin{itemize}
    \item We can reproduce the size, shape, and spatial orientation of Sgr A East with a supernova exploding in a simple two-component medium consisting of a spherical molecular cloud and a (spherically symmetric) wind from the young nuclear star cluster.
    \begin{itemize}
    \item[$\bullet$] This two-component scenario restricts the position of the SN explosion quite sharply.
    It also places constraints on the position of the molecular cloud, although there is more freedom in placing the cloud than the supernova. The two-component scenario shows the best position of the 
    SN explosion to be at a total distance of 3.7 pc from \SgrAstar, located 3 pc behind it and slightly eastwards, and 2 pc below the Galactic plane. 
    The cloud should be 10 pc below the plane, 2 pc eastwards from the Galactic centre, and 2 pc behind \SgrAstar.
    \item[$\bullet$] The simulation that best mimics Sgr A East has an age of 10 kyr, but there are
    other models of reasonably high fitness  with ages of 19 and 29 kyr. However, we were unable to find good models with ages younger than 5 kyr. Therefore, from our simulations, we cannot place an upper limit on the age, but obtain a rather strict lower limit of 5 kyr. 
    \item[$\bullet$] The age of our best model, 10 kyr, is consistent with a possible connection between Sgr A East and the Cannonball pulsar wind nebula. 
    \item[$\bullet$] The projected location of the warm dust patch or cloudlet suggested by \citet{lau2015} to be in the same direction 
    from the Galactic centre as the initial position of the best-fit supernova model, but farther away, $\sim5$ pc from \SgrAstar. The $\sim$1 - 1.5 pc difference in the $x-$ and $z$ -coordinates 
    is not consistent with the best-fit model, nor does the supernova explosion in the position of the cloud give good results. However, based on the projected positions of the supernova explosion, the dust cloud, and the Cannonball, we cannot exclude that they are directly linked if the cloud shared the initial velocity of the Sgr A East progenitor.
    \item[$\bullet$] The feeding rate of matter into the central parsec in the best models of our ISM scenario (wind+cloud) is lower than $10^{-3}\MSun \mathrm{yr}^{-1}$ at the present time and well into the future, but we predict an increased feeding rate in about 70 kyr.
    \end{itemize}
    \item Alternatively, we find some not-so-good-but-acceptable models for the ISM
    consisting of one molecular cloud alone. As before, the position of the SN explosion 
    is restricted more than the position of the cloud. We also find solutions with
    differing ages, with fitness maxima at 5, 12, and 20 kyr. Again, we cannot find a good solution
    for very young ages. In this case, the SN explosion is located slightly (1 pc) below the plane,
    3 pc eastwards, and at about the same line-of-sight distance as \SgrAstar.
    The cloud lies 10 pc eastwards from \SgrAstar, at the same line-of-sight distance and slightly below the plane.
    \item The ISM consisting only of the wind from the nuclear star cluster does not
    produce acceptable models for Sgr A East.
    \item Based on simulations presented in previous papers, it is also possible to simulate Sgr A East as an explosion in a homogeneous medium (with a density around
    1000 cm$^{-3}$) and with the remnant distorted by differential rotation. However, the time needed to achieve the proper shape is longer than 30 kyr. 
\end{itemize}

Our best model is quite successful in predicting the shape
of Sgr A East supernova remnant, but there are possibilities to 
improve it further. The molecular cloud could be treated in a more
sophisticated way, in tune with continuously improving observations by ALMA and other radiotelescopes. Then, even though we do not assume the circumnuclear disk to play a significant role in the evolution of Sgr A East, it might be an important factor in some external galaxies, which we plan to explore in connection with the mass delivery to the central regions of galaxies.

\section*{Acknowledgments}

This study has been supported by Czech Science Foundation Grant 19-15480S and by the project RVO 67985815. 
BB is supported the by the \'UNKP-22-4-SZTE-476 New National Excellence Program of the Ministry for Culture and Innovation from the source of the National Research, Development and Innovation Fund, and the Hungarian NKFIH/OTKA FK-134432 grant.
Authors are grateful to Guillermo Tenorio-Tagle for inspiring discussions and
constructive comments. 

\bibliographystyle{aa} 
\bibliography{sgraest_ref} 

\begin{appendix}
\section{Fitness function}
\label{A1}
\begin{figure}
  \includegraphics[angle=0,width=0.45\textwidth]{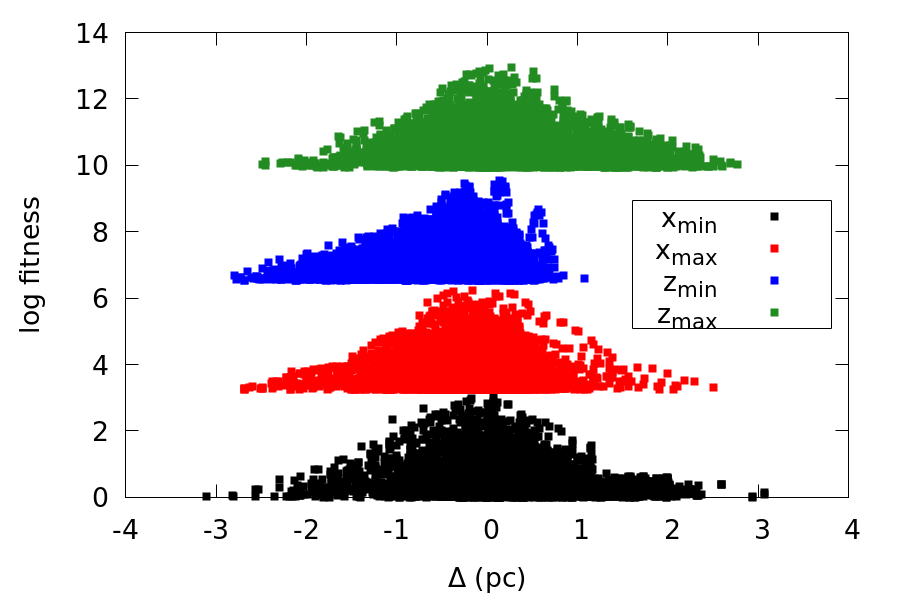}
  
  \includegraphics[angle=0,width=0.45\textwidth]{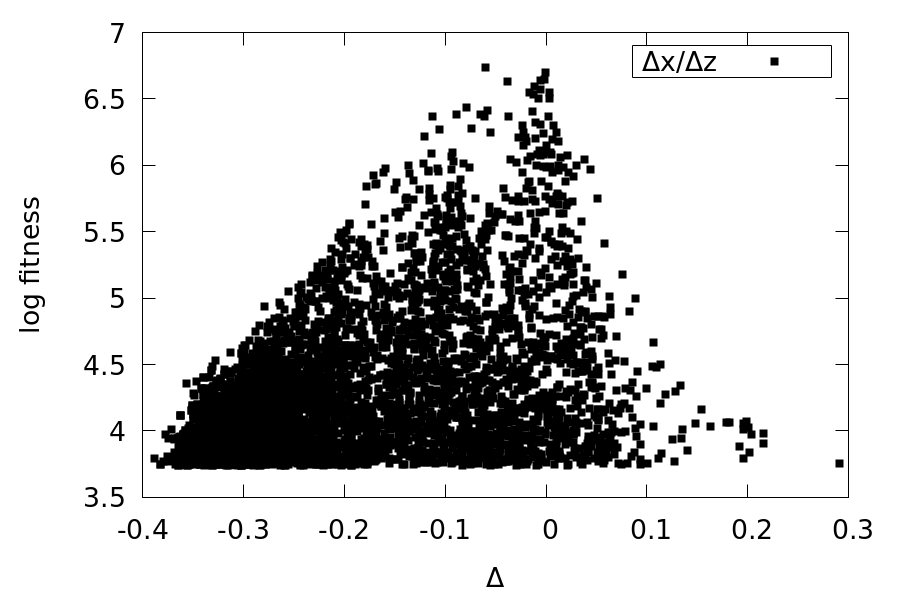}
  
  \includegraphics[angle=0,width=0.45\textwidth]{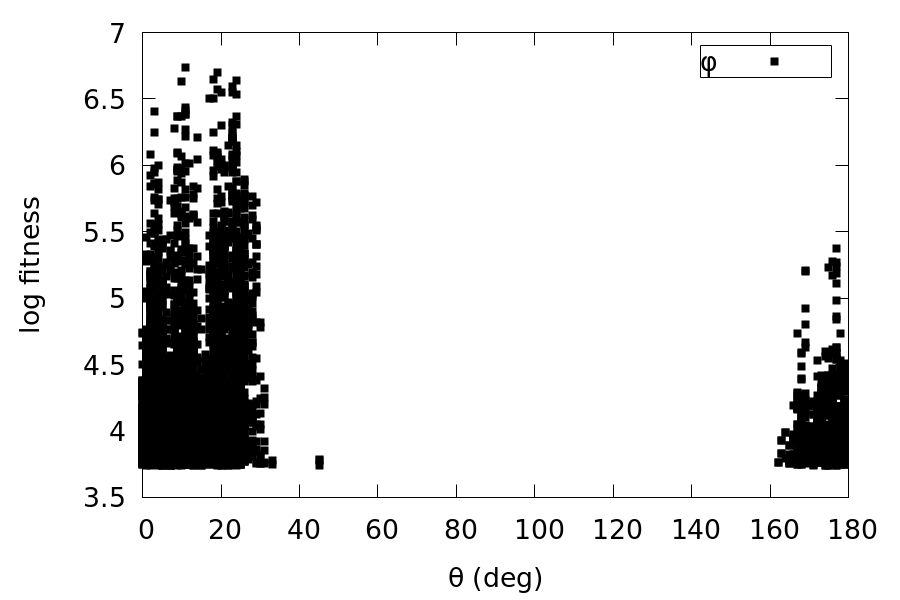}
      \caption{Fitness for models in the wind+cloud scenario and dependence on
      the parameters we used. Only models with a fitness $f \ge 0.01f_{\rm best}$ are
      shown. The upper panel shows the distribution of 
      $x_{\rm min}-x_{\rm min}^{\rm obs}$ (black),
      $x_{\rm max}-x_{\rm max}^{\rm obs}$ (red), 
      $z_{\rm min}-z_{\rm min}^{\rm obs}$ (blue),
      and $z_{\rm max}-z_{\rm max}^{\rm obs}$ (green);
      the desired value is set on 0. The second row shows the distribution
      for the ratio $\frac{x_{\rm max}-x_{\rm min}}{z_{\rm max}-z_{\rm min}}$; $\Delta$ is 
      the difference between this ratio and the desired value of 1.38.
      The bottom row shows the distribution of the angle $\phi$ between the main axis of the structure
      and the $x-$coordinate (i.e. the Galactic plane), the desired value is $0^{\circ}$ or $180^{\circ}$.
      These quantities enter the fitness function directly.
      Values of fitness for different quantities are artificially offset.
              }
         \label{fitness_parameters}
\end{figure}

\begin{figure}
   \includegraphics[angle=0,width=0.45\textwidth]{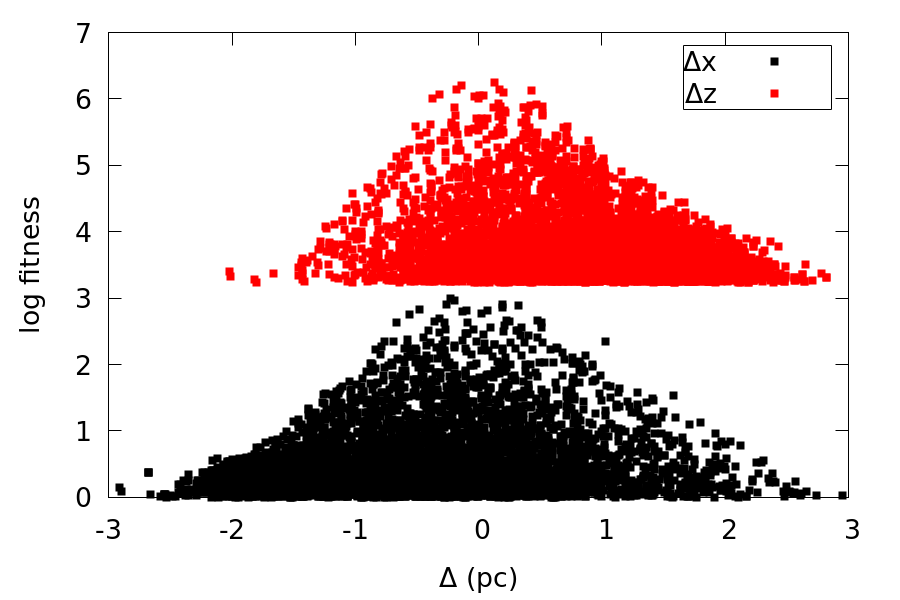}
      \caption{Fitness for models in the wind+cloud scenario and sizes 
      $\Delta x = x_{\rm max} - x_{\rm min}$ and 
      $\Delta z = z_{\rm max} - z_{\rm min}$.
      Only models with the fitness $f \ge 0.01f_{\rm best}$ are
      shown. These quantities are not directly involved in the fitness function.
      The graph shows $\Delta x-\Delta x_{\rm obs}$ (black) and
      $\Delta z-\Delta z_{\rm obs}$ (red). Values of fitness for different quantities
      are artificially offset.
              }
         \label{fitness_dxdz}
\end{figure}

Here we give a more detailed description of the fitness function we used to select the
best (or good enough) models of Sgr A East. Its basic description is given
in Sect. \ref{subsec:fitness} (Fitness function).

The fitness function has six input parameters: $x_{\rm min}, x_{\rm max}$, 
$z_{\rm min}$, $z_{\rm max}$, $\phi$, and $\frac{\Delta x}{\Delta z}$.
The first four, $x_{\rm min}$, $x_{\rm max}$, $z_{\rm min}$,  and $z_{\rm max}$ , are 
minimum and maximum positions reached by a studied model at a given age, 

\begin{equation}
\begin{split}
f = 
w_{xmin}\left(\frac{x_{\rm min}-x_{\rm min}^{\rm obs}}{\sigma_{xz}}\right)^2 
+ w_{xmax}\left(\frac{x_{\rm max}-x_{\rm max}^{\rm obs}}{\sigma_{xz}}\right)^2 \\
+ w_{zmin}\left(\frac{z_{\rm min}-z_{\rm min}^{\rm obs}}{\sigma_{xz}}\right)^2
+ w_{zmax}\left(\frac{z_{\rm max}-z_{\rm max}^{\rm obs}}{\sigma_{xz}}\right)^2 \\
+ w_{\phi}\left(\frac{\phi -\phi^{\rm obs}}{\sigma_\phi}\right)^2 \\
+ w_{\frac{\Delta x}{\Delta z}}\left(\frac{\frac{\Delta x}{\Delta z} - \frac{\Delta x}{\Delta z}^{\rm obs}}{\sigma_\frac{\Delta x}{\Delta z}}\right)^2 
\end{split}
.\end{equation}
Because input parameters do not have the same units, we have to normalize them. We adopted the following values: $\sigma_{xz}$ = 1 pc, $\sigma_\phi$ = 30$^\circ$, and $\sigma_\frac{\Delta x}{\Delta z}$ = 0.2. Weights $w$ in the above formula used in our analysis were 1.0 except $w_\frac{\Delta x}{\Delta z}$, which was equal to 2.0. Different weights were tested with similar results. We also tried to use the size ($\Delta x$ and $\Delta z$) as input parameters,
but the results were better when only positions $x_{\rm min}$, $x_{\rm max}$,
$z_{\rm min}$, and $z_{\rm max}$ were used.

Figure \ref{fitness_parameters} shows the fitness distribution for all six input parameters
for the calculations in the wind+cloud scenario (only models with the fitness  $f \ge 0.01f_{\rm best}$ 
are shown). The best models tend to have the desired values.
The fitness as a function of the size (Fig. \ref{fitness_dxdz}) also favours the
observed (desired) value, although the size is not an input parameter in the fitness function.
The distribution is not always symmetric around the desired value, which is connected to the fact 
that many models are more spherical than Sgr A East.

\section{Effects of the homogeneous background density on the best model}
\label{A2}

\begin{figure}
   \includegraphics[angle=0,width=0.45\textwidth]{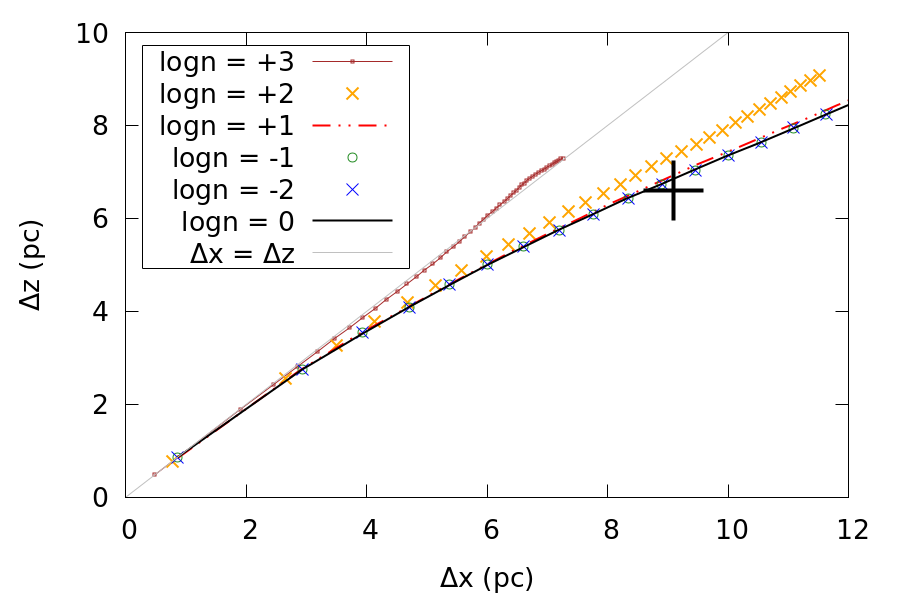}
      \caption{Evolution of dimensions ($\Delta x$, $\Delta z$) of the supernova 
      remnant evolving into the ISM composed of the molecular cloud (with the
      position and size derived by the model A) and the young nuclear cluster wind,
      with the homogeneous
      background with a varying density. The solid black line corresponds to the
      density of $1\ \mbox{cm}^{-3}$ (the standard case). Green and blue
      symbols correspond to lower densities (0.1 and 0.01 $\mbox{cm}^{-3}$, respectively).
      The dashed red line and yellow crosses belong to higher densities (10 and 100 $\mbox{cm}^{-3}$).
      The brown line with squares corresponds to the case of 1000 $\mbox{cm}^{-3}$. 
      Symbols, if present, are plotted each 1 kyr.
      The grey line is $\Delta x = \Delta z$, and the black cross shows the observed dimensions
      of Sgr A East.
              }
         \label{back_dxdz}
\end{figure}

Our ISM model contains two principal components: the molecular cloud M50, and the wind from
the young cluster, and then a low-density homogeneous background. This background
has a negligible influence on the evolution of the supernova remnant because the density
is dominated by the principal components. To examine this limited influence, we
performed a set of simulations in the ISM (the molecular cloud M50 and the
young nuclear cluster wind)
according to our best model (model A), but with different values
of the homogeneous background density. Calculations were made with an external
gravitational field, but the results are essentially the same without it (see Sect. \ref{subsec:grav}).

Fig. \ref{back_dxdz} shows $x$ and $z$ sizes calculated for different values of the homogeneous
background density and compares them to our standard case (model A). It follows from the figure 
that the effects of background are small or negligible
for values of the background density below $10^2\ \mbox{cm}^{-3}$. 
Calculations with a density of $1\ \mbox{cm}^{-3}$ (standard) and below are indistinguishable from each
other, and calculations with the density of $10\ \mbox{cm}^{-3}$ are just marginally different.

Calculations in a density of $100\ \mbox{cm}^{-3}$ are influenced by the increased density.
The evolution proceeds slightly more slowly, as visible in Fig. \ref{back_dxdz}; the symbols
are plotted each 1 kyr. To reach the observed $x$-size of Sgr A East (9.1 pc), the remnant has to
evolve for 17 kyr (only 10 kyr in a lower density background) 
and its $z$-size at that time is larger than the observed value (i.e. the remnant is more
spherical because the density gradients in the ISM are smaller due to the increased background), but
results are still comparable to our standard case.

Results with a density of $10^3\ \mbox{cm}^{-3}$ are different from our standard case and
incompatible with observations of Sgr A East: The remnant grows too slowly and does not 
have the proper elongation (the last point in Fig. \ref{back_dxdz} corresponds to an
age of 50 kyr). 
In this case, the background density is an important, not negligible
part of the ISM distribution, and in many aspects, this scenario resembles the evolution
of remnants in the homogeneous high-density distribution (Section \ref{subsec:hmg}).

It is difficult to estimate the value that this background could reasonable have. For the central molecular zone, which is outside our computational box, \citet{2019ApJ...883...54O} concluded that its volume is dominated by warm and diffuse gas with an average density of about $50\ \mbox{cm}^{-3}$ and a temperature of about 200 K. This is still in the region in which our results for the ISM dominated by the cloud and wind apply, but we assume that the density outside the molecular zone is much lower.

\end{appendix}

\end{document}